# Efficient Computation of the Shapley Value
# for Game-Theoretic Network Centrality

**Tomasz P. Michalak**　　　　　　　　　　　　Tomasz.Michalak@cs.ox.ac.uk
*Department of Computer Science, University of Oxford*
*OX1 3QD, UK*
*Institute of Informatics, University of Warsaw*
*02-097 Warsaw, Poland*

**Karthik .V. Aadithya**　　　　　　　　　　　　Aadithya@eecs.berkeley.edu
*Department of Electrical Engineering and Computer Sciences*
*University of California*
*Berkeley, CA 94720-4505, United States, USA*

**Piotr L. Szczepański**　　　　　　　　　　P.Szczepanski@stud.elka.pw.edu.pl
*Institute of Informatics*
*Warsaw University of Technology*
*00-661 Warsaw, Poland*

**Balaraman Ravindran**　　　　　　　　　　　　Ravi@cse.iitm.ac.in
*Computer Science and Engineering*
*Indian Institute of Technology Madras*
*Chennai, 600 036, India*

**Nicholas R. Jennings**　　　　　　　　　　　　NRJ@ecs.soton.ac.uk
*School of Electronics and Computer Science*
*University of Southampton*
*SO17 1BJ Southampton, UK*

## Abstract

The Shapley value—probably the most important normative payoff division scheme in coalitional games—has recently been advocated as a useful measure of centrality in networks. However, although this approach has a variety of real-world applications (including social and organisational networks, biological networks and communication networks), its computational properties have not been widely studied. To date, the only practicable approach to compute Shapley value-based centrality has been via Monte Carlo simulations which are computationally expensive and not guaranteed to give an exact answer. Against this background, this paper presents the first study of the computational aspects of the Shapley value for network centralities. Specifically, we develop exact analytical formulae for Shapley value-based centrality in both weighted and unweighted networks and develop efficient (polynomial time) and exact algorithms based on them. We empirically evaluate these algorithms on two real-life examples (an infrastructure network representing the topology of the Western States Power Grid and a collaboration network from the field of astrophysics) and demonstrate that they deliver significant speedups over the Monte Carlo approach. For instance, in the case of unweighted networks our algorithms are able to return the exact solution about 1600 times faster than the Monte Carlo approximation, even if we allow for a generous 10% error margin for the latter method.

## 1. Introduction

In many network applications, it is important to determine which nodes and edges are more critical than others. Classic examples include identifying the most important hubs in a road network (Schultes & Sanders, 2007), the most critical functional entities in a protein network (Jeong, Mason, Barabasi, & Oltvai, 2001), or the most influential people in a social network (Kempe, Kleinberg, & Tardos, 2003). Consequently, the concept of *centrality*, which aims to quantify the importance of individual nodes/edges, has been extensively studied in the literature (Koschützki, Lehmann, Peeters, Richter, Tenfelde-Podehl, & Zlotowski, 2005; Brandes & Erlebach, 2005).

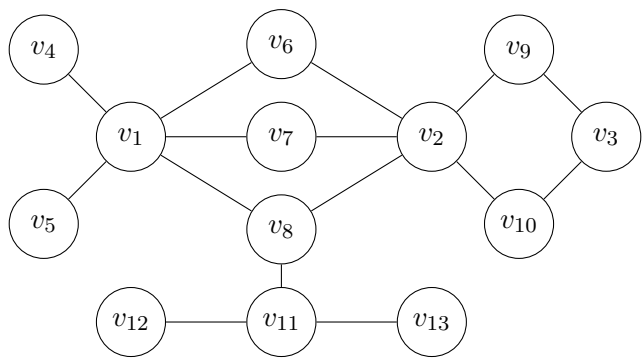

Figure 1: Sample network of 13 nodes

Generally speaking, centrality analysis aims to create a *consistent ranking* of nodes within a network. To this end, *centrality measures* assign a score to each node that in some way corresponds to the importance of that node given a particular application. Since "importance" depends on the context of the problem at hand, many different centrality measures have been developed. Three of the most well-known and widely applied are: *degree centrality*, *closeness centrality* and *betweenness centrality*.[1] In this paper, we refer to these measures as *conventional/standard centrality*. Degree centrality, in brief, quantifies the power of a node by its degree, i.e., by the number of its adjacent edges. For instance, in the sample network in Figure 1, nodes $v_1$ and $v_2$ have degree 5 and, if judged by degree centrality, these are the most important nodes within the entire network. Conversely, closeness centrality focuses on distances among nodes and gives high value to the nodes that are close to all other nodes. With this measure, node $v_8$ in Figure 1 is ranked top. The last measure—betweenness centrality—considers shortest paths (i.e., paths that use the minimal number of links) between any two nodes in the network. The more shortest paths the node belongs to, the more important it is. With this measure, $v_2$ in Figure 1 is more important than all the other nodes (including $v_1$ and $v_8$, which are chosen by other measures as the most important node). Clearly, all these measures expose different characteristics of a node. Consider, for instance, an epidemiology application, where the aim is to identify those people (i.e., nodes) in the social network who have the biggest influence on the spread of the disease and should become a focal point of any prevention or emergency measures. Here, degree centrality ranks top nodes with the biggest immediate *sphere of influence*—their infection would lead to the highest number of adjacent nodes being exposed to the disease. On the other hand, closeness centrality identifies those nodes whose infection would lead to the *fastest* spread of the disease throughout the society. Finally, betweenness centrality reveals the nodes that play a crucial role in *passing* the disease from one person in a network to another.[2]

---

1. Koschützki et al. (2005) and Brandes and Erlebach (2005) give a good overview of these and other centrality measures.

The common feature of all the aforementioned standard measures is that they assess the importance of a node by focusing only on the role that a node plays by itself. However, in many applications such an approach is inadequate because of *synergies* that may occur if the functioning of nodes is considered in groups. Referring again to Figure 1 and our epidemiology example, a vaccination of individual node $v_6$ (or $v_7$ or $v_8$) would not prevent the spread of the disease from the left to the right part of the network (or vice versa). However, the simultaneous vaccination of $v_6$, $v_7$ and $v_8$ would achieve this goal. Thus, in this particular context, nodes $v_6$, $v_7$ and $v_8$ do not play any significant role individually, but together they do. To quantify the importance of such groups of nodes, the notion of *group centrality* was introduced by Everett and Borgatti (1999). Intuitively, group centrality works broadly the same way as standard centrality, but now the focus is on the functioning of a given group of nodes, rather than individual nodes. For instance, in Figure 1, the group degree centrality of $\{v_1, v_2\}$ is 7 as they both have 7 distinct adjacent nodes.

Although the concept of group centrality addresses the issue of synergy between the functions is played by various nodes, it suffers from a fundamental deficiency. It focuses on particular, a priori determined, groups of nodes and it is not clear how to construct a consistent ranking of *individual* nodes using such group results. Specifically, should the nodes from the most valuable group be ranked top? Or should the most important nodes be those which belong to the group with the highest average value per node? Or should we focus on the nodes which contribute most to every coalition they join? In fact, there are very many possibilities to choose from.

A framework that does address this issue is the *game theoretic network centrality measure*. In more detail, it allows the consistent ranking of individual nodes to be computed in a way that accounts for various possible synergies occurring within possible groups of nodes (Grofman & Owen, 1982; Gómez, González-Arangüena, Manuel, Owen, Del Pozo, & Tejada, 2003; Suri & Narahari, 2008). Specifically, the concept builds upon cooperative game theory—a part of game theory in which agents (or players) are allowed to form coalitions in order to increase their payoffs in the game. Now, one of the fundamental questions in cooperative game theory is how to distribute the surplus achieved by cooperation among the agents. To this end, Shapley (1953) proposed to remunerate agents with payoffs that correspond to their individual marginal contributions to the game. In more detail, for a given agent, such an individual marginal contribution is measured as the weighted average marginal increase in the payoff of any coalition that this agent could potentially join. Shapley famously proved that his concept—known since then as the Shapley value—is the only division scheme that meets certain desirable normative properties.[3] Given this, the key idea of the game theoretic network centrality is *to define a cooperative game over a network* in which agents are the nodes, coalitions are the groups of nodes, and payoffs of coalitions are defined so as to meet requirements of a given application. This means that the Shapley value of each agent in such a game can then be interpreted as a *centrality measure* because it represents *the average marginal contribution made by each node to every coalition of the other nodes*.[4] In other words, the Shapley value answers the question of how to construct a consistent ranking of individual nodes once groups of nodes have been evaluated.

---

2. For the differences in interpretation of standard centralities see the work of Borgatti and Everett (2006).
3. See Section 3 for more details.

In more detail, the Shapley value-based approach to centrality is, on one hand, much more *sophisticated* than the conventional measures, as it accounts for any group of nodes from which the Shapley value derives a consistent ranking of individual nodes. On the other hand, it confers a high degree of *flexibility* as the cooperative game over a network can be defined in a variety of ways. This means that *many different versions of Shapley value-based centrality* can be developed depending on the particular application under consideration, as well as on the features of the network to be analyzed. As a prominent example, in which a specific Shapley value-based centrality measure is developed that is crafted to a particular application, consider the work of Suri and Narahari (2008) who study the problem of selecting the top-$k$ nodes in a social network. This problem is relevant in all those applications where the key issue is to choose a group of nodes that together have the biggest *influence* on the entire network. These include, for example, the analysis of co-authorship networks, the diffusion of information, and viral marketing. As a new approach to this problem, Suri and Narahari define a cooperative game in which the value of any group of nodes is equal to the number of nodes within, and adjacent to, the group. In other words, it is assumed that the agents' *sphere of influence* reaches the immediate neighbors of the group. Whereas the definition of the game is a natural extension of the (group) degree centrality discussed above, the *Shapley value of nodes in this game* constitutes a new centrality metric that is, arguably, qualitatively better than standard degree centrality as far as the node's influence is concerned. The intuition behind it is visible even in our small network in Figure 1. In terms of influence, node $v_1$ is more important than $v_2$, because it is the only node that is connected to $v_4$ and $v_5$. Without $v_1$ it is impossible to influence $v_4$ and $v_5$, while each neighbor of $v_2$ is accessible from some other node. Thus, unlike standard degree centrality, which evaluates $v_1$ and $v_2$ equally, the centrality based on the Shapley value of the game defined by Suri and Narahari recognizes this difference in influence and assigns a higher value to $v_1$ than to $v_2$.

Unfortunately, despite the advantages of Shapley value-based centrality over conventional approaches, efficient algorithms to compute it have not yet been developed. Indeed, given a network $G(V, E)$, where $V$ is the set of nodes and $E$ the set of edges, using the original Shapley value formula involves computing the marginal contribution of every node to every coalition which is $O(2^{|V|})$. Such an exponential computation is clearly prohibitive for bigger networks (of, e.g. 100 or 1000 nodes). For such networks, the only feasible approach currently outlined in the literature is Monte-Carlo sampling (e.g., Suri & Narahari, 2008; Castro, Gomez, & Tejada, 2009). However, this method is not only inexact, but can be also very time-consuming. For instance, as shown in our simulations, for a weighted network of about 16,000 nodes and about 120,000 edges, the Monte Carlo approach has to iterate 300,000 times through the entire network to produce the approximation of the Shapley value with a 40% error margin.[5] Moreover, exponentially more iterations are needed to further reduce this error margin.

---

4. We note that other division schemes or power indices from cooperative game theory, such as Banzhaf power index (Banzhaf, 1965), could also be used as centrality measures (see, for instance, the discussion in the work by Grofman & Owen, 1982). However, like most of the literature, we focus on the Shapley value due to its desirable properties.

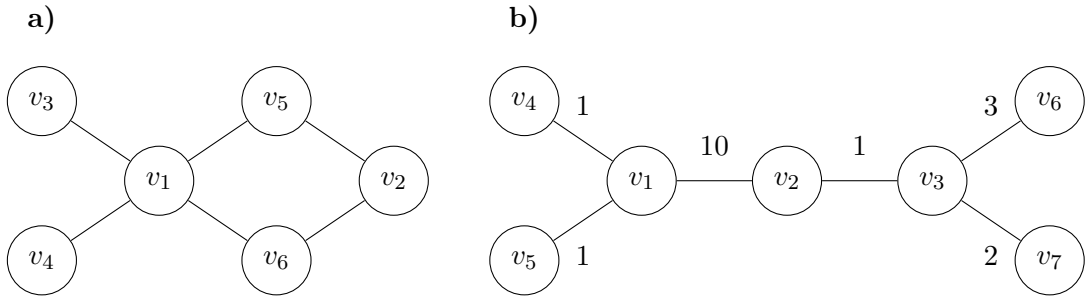

Figure 2: Sample unweighted and weighted networks of 6 and 7 nodes, respectively.

Against this background, we develop *polynomial-time algorithms to compute Shapley value-based centrality. Specifically, we focus on five underlying games* defined over a network; those games extend, in various directions, standard notions of degree and closeness centrality. As our starting point, we consider the game defined by Suri and Narahari and propose an exact, linear-time algorithm to compute the corresponding Shapley value-based centrality. We denote this game by $g_1$. We then analyse the computational properties of four other games defined over networks. We denote them $g_2, g_3, g_4$, and $g_5$, respectively. While each of these games captures a different *flavor of centrality*, they all, similarly to the game of Suri and Narahari, embrace one fundamental centrality idea: given a group of nodes $C$, the function that defines the value of $C$ in the game must somehow quantify the *sphere of influence* of $C$ over other nodes in the network. In particular:

$g_2$   In this game the value of coalition $C$ is a function of its own size and of the number of nodes that are immediately reachable in at least $k$ different ways from $C$. This game is inspired by Bikhchandani, Hirshleifer, and Welch (1992) and is an instance of the general threshold model introduced by Kempe, Kleinberg, and Tardos (2005). It has a natural interpretation: an agent "becomes influenced" (with ideas, information, marketing message, etc.) only if at least $k$ of his neighbors have already become influenced. For instance, given $k = 2$, the value of coalition $\{v_1, v_2\}$ in Figure 2a is 4 as the coalition is of size 2 and there are two neighbors with no less than 2 edges adjacent to this coalition.

$g_3$   This game concerns weighted graphs (unlike $g_1$ and $g_2$). Here, the value of coalition $C$ depends on its size and on the set of all nodes within a cutoff distance of $C$, as measured by the shortest path lengths on the weighted graph. For example, in Figure 2b, if the cutoff is set to 8 then coalition $\{v_2\}$ has value 4 as it is able to influence 3 nodes $v_3$, $v_6$, and $v_7$ that are not further than 8 away from $\{v_2\}$. The cutoff distance should be interpreted here as a "radius" of the sphere of influence of any coalition.

---

5. See Section 5 for the exact definition of the error margin.

| Game | Graph | Value of a coalition $C$, i.e., $\nu(C)$ | Complexity | Accuracy |
|------|-------|------------------------------------------|------------|----------|
| $g_1$ | $UW$ | $\nu(C)$ is the number of nodes in $C$ and those immediately reachable from $C$ | $O(|V| + |E|)$ | exact |
| $g_2$ | $UW$ | $\nu(C)$ is the number of nodes in $C$ and those immediately reachable from $C$, but via at least $k$ different edges | $O(|V| + |E|)$ | exact |
| $g_3$ | $W$ | $\nu(C)$ is the number of nodes in $C$ and those not further than $d_{cutoff}$ away | $O(|V||E| + |V|^2 \log|V|)$ | exact |
| $g_4$ | $W$ | $\nu(C)$ is the sum of $f(.)$'s — the non--increasing functions of the distance between $C$ and other nodes | $O(|V||E| + |V|^2 \log|V|)$ | exact |
| $g_5$ | $W$ | $\nu(C)$ is the number of nodes in $C$ and those directly connected to $C$ via edges which sum of weights exceeds $W_{cutoff}$ | $O(|V||E|)$ | approx. $\sim$ 5-10% |

Table 1: Games considered in this paper and our results ($UW$ denotes unweighted graphs and $W$ weighted).

$g_4$ This game generalizes $g_3$ by allowing the value of $C$ to be specified by its size and an arbitrary non-increasing function $f(.)$ of the distance between $C$ and the other nodes in the network. For instance, the value of $\{v_1, v_3\}$ in Figure 2b when our function is $f(d) = \frac{1}{1+d}$ is $2 \times 1 + 3 \times \frac{1}{2} + 1 \times \frac{1}{3} + 1 \times \frac{1}{4} = 4\frac{1}{12}$. The intuition here is that the coalition has more influence on closer nodes than on those further away—a property that cannot be expressed with the standard closeness centrality. Thus, $g_4$ can be seen as an extension of closeness centrality.

$g_5$ The last game is an extension of $g_2$ to the case of weighted networks. Here, the value of $C$ depends on the adjacent nodes that are connected to the coalition with weighted edges whose sum exceeds a given threshold $w_{cutoff}$ (recall that in $g_2$ this threshold is defined simply by the integer $k$). Whereas in $g_3$ and $g_4$ weights on edges are interpreted as distance, in $g_5$ they should be interpreted as a *power of influence*. For example, in Figure 2b, when the threshold for each vertex is 5, the value of coalition $\{v_1, v_3\}$ is 3 because this coalition of size two has only enough power to influence one additional node $v_2$.

The computation of the Shapley value for each of the above five games (see Table 1 for an overview) will be the main focus of the paper. These Shapley values are extensions of either degree or closeness centrality metrics and their applications are all those settings in which the influence of nodes on other nodes in the network has to be evaluated. Our results can be summarized as follows:

- We demonstrate that it is possible to *exactly and efficiently compute a number of Shapley value-based network centrality measures.* Our methods take advantage of both the network structure, as well as the specifics of the underlying game defined over a network.

- For the first four games, we derive closed-form expressions for the Shapley values. Based on these, we provide *exact linear and polynomial-time algorithms* that efficiently compute the Shapley values, i.e., without the need to enumerate all possible coalitions. Specifically, our algorithms run in $O(|V| + |E|)$ for $g_1$ and $g_2$ and in $O(|V||E| + |V|^2\log|V|)$ for $g_3$ and $g_4$. Furthermore, for the fifth measure of centrality, we develop *a closed-form polynomial time computable Shapley value approximation*. This algorithm has running time $O(|V||E|)$ and our experiments show that its approximation error is about 5% for large networks. The summary of our algorithms' performance can be found in Table 1.

- We evaluate our algorithms on *two real-life examples*: an infrastructure network representing the topology of the Western States Power Grid and a collaboration network from the field of astrophysics. The results show that our algorithms deliver significant speedups over Monte Carlo simulations. For instance, given the unweighted network of Western States Power Grid, our algorithms return the exact Shapley value for $g_1$ and $g_2$ about 1600 times faster than the Monte Carlo method returns an approximation with the 10% error margin.

The remainder of the paper is organized as follows. In Section 2 we discuss related work. Notation and preliminary definitions are presented in Section 3. In Section 4 we analyse the five types of centrality-related coalitional games and propose polynomial time Shapley value algorithms for all of them. The results of numerical simulations are presented in Section 5 (with some details on the simulation setup presented in Appendix A). Conclusions and future work follow. Finally, Appendix B provides a summary of the key notational conventions.

## 2. Related Literature

The issue of centrality is one of the fundamental research directions in the network analysis literature. In particular, Freeman (1979) was the first to formalise the notion of centrality by presenting the conventional centrality measures: degree, closeness and betweenness. Many authors have subsequently worked on developing new centrality measures, or refining existing ones (e.g., Bonacich, 1972; Noh & Rieger, 2004; Stephenson & Zelen, 1989), and developing algorithms for efficient centrality computation (e.g., Brandes, 2001; Eppstein & Wang, 2001). In this context, Grofman and Owen (1982) were the first to apply game theory to the topic of centrality, where they focused on the Banzhaf power index (Banzhaf, 1965). In a follow-up work, Gómez et al. (2003) combined Myerson's (1977) idea of graph-restricted games (in which each feasible coalition is induced by a subgraph of a graph) with the concept of centrality and proposed new Shapley value-based network centrality measures. In contrast to Gómez et al., Suri and Narahari (2008, 2010) assumed all coalitions to be feasible, which is the approach we also adopt in this paper.

The fundamental problem with the conventional models of coalitional games, i.e., their *exponential complexity in the number of agents*, that we tackle in this paper, has been also studied in the literature on algorithmic aspects of coalitional games. Indeed, since the seminal work of Deng and Papadimitriou (1994), this issue has received considerable attention from computer scientists. Specifically, as an alternative to the straightforward (but exponential) listing of all possible coalitions, a number of authors have proposed more efficient *representations* for coalitional games. Such representations fall into two main categories (Wooldridge & Dunne, 2006):

- Those that give the characteristic function a specific interpretation in terms of combinatorial structures such as graphs. This is the approach adopted by, for instance, Deng and Papadimitriou (1994), Greco, Malizia, Palopoli, and Scarcello (2009), and Wooldridge and Dunne (2006) and its advantage is that the ensuing representation is always guaranteed to be succinct. However, the disadvantage is that it is not always fully expressive, i.e., it cannot represent all coalitional games.

- Those that try to find a succinct, but still fully expressive, representation. This is, for instance, the approach adopted by Conitzer and Sandholm (2004), Ieong and Shoham (2005), and Elkind, Goldberg, Goldberg, and Wooldridge (2009). These representations are more general in that they completely capture all coalitional games of interest, although they are not always guaranteed to be succinct.

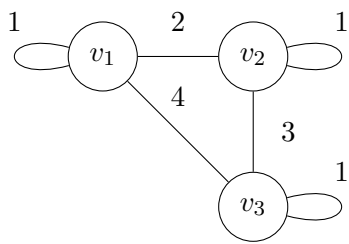

Figure 3: The Induced-subgraph representation of a sample coalitional game of 3 players. In this game, values of coalitions $\{v_1\}$, $\{v_2\}$, $\{v_3\}$, $\{v_1, v_2\}$, $\{v_1, v_3\}$, $\{v_2, v_3\}$, and $\{v_1, v_2, v3\}$ are 1, 1, 1, 1+1+2, 1+1+3, 1+1+4, and 1+1+1+2+3+4, respectively.

Unfortunately, even for some succinctly representable games, computing the Shapley value has been shown to be NP-Hard (or even worse, #P-Complete) for many domains, including weighted voting games (Deng & Papadimitriou, 1994), threshold network flow games (Bachrach & Rosenschein, 2009) and minimum spanning tree games (Nagamochi, Zeng, Kabutoya, & Ibaraki, 1997). Similarly, Aziz, Lachish, Paterson, and Savani (2009a) obtained negative results for a related problem of computing the Shapley-Shubik power index for the spanning connectivity games that are based on undirected, unweighted multigraphs. Also, Bachrach, Rosenschein, and Porat (2008b) showed that the computation of the Banzhaf index for connectivity games, in which agents own vertices and control adjacent edges and aim to become connected to the certain set of primary edges, is #P-Complete.

Fortunately, some positive results have also been discovered. Probably the most known among these are due to Deng and Papadimitriou (1994) and Ieong and Shoham (2005). In more detail, Deng and Papadimitriou proposed a representation based on weighted graphs, where a node is interpreted as an agent, and the weight of an edge is interpreted as the value of cooperation between the two agents that are connected by this edge.[6] The value of any coalition is then defined as the sum of weights of all its internal edges, or, in other words, the weights of edges belonging to a subgraph induced by members of the coalition. A three-player example of this formalism, called the *induced-subgraph representation*, can be found in Figure 3. The downside of this representation is that it is not fully expressive. However, the upside is that, for games that can be formalised as weighted graphs, this representation is always concise. Furthermore, it allows the Shapley value to be computed in time linear in the number of players. Specifically, in this case, the Shapley value is given by the the following formula:

$$Shapley\ Value(v_i)\ = v_i\text{'s self-loop weight} + \sum_{v_j \in \text{neighbours of } v_i} \frac{\text{weight of edge from } v_i \text{ to } v_j}{2}. \quad (1)$$

Ieong and Shoham (2005) developed a representation consisting of a finite set of logical rules of the following form: *Boolean Expression* → *Real Number*, with agents being the atomic boolean variables. In this representation, the value of a coalition is equal to the sum of the right sides of those rules whose left sides are satisfied by the coalition. This representation, called *marginal contribution networks* (or MC-Nets for short) is (i) fully expressive (i.e., it can be used to model any game), (ii) exponentially more concise for some games, and most importantly, (iii) allows the Shapley value to be computed in time linear in the size of the representation, provided the boolean expressions in all rules are conjunctions of (either positive or negative) atomic literals. In MC-Nets, the rules have an interesting game-theoretic interpretation, as each rule directly specifies an incremental marginal contribution made by the agents featured in that rule. Now, using the additivity axiom met by the Shapley value, it is possible to consider every rule as a separate "simple" game, then using other axioms straightforwardly compute the Shapley value for this "simple" game, and, finally, sum up the results for all "simple" games to obtained the Shapley value. Building on this, Elkind et al. (2009) developed extensions of MC-Nets to more sophisticated (read-once) boolean expressions, while Michalak, Marciniak, Samotulski, Rahwan, McBurney, Wooldridge, and Jennings (2010a), Michalak, Rahwan, Marciniak, Szamotulski, and Jennings (2010b) developed generalizations to coalitional games with externalities. Another recently proposed representation formalism for coalitional games that allows for polynomial calculations of the Shapley value are decision diagrams (Bolus, 2011; Aadithya, Michalak, & Jennings, 2011; Sakurai, Ueda, Iwasaki, Minato, & Yokoo, 2011). Now, while MC-Nets offer a fully-expressive representation that works for arbitrary coalitional games, it is possible to speed up the Shapley value computation by focusing on specific (not necessarily fully expressive) classes of games. One particular class of games that has been investigated in detail is weighted voting, for which both approximate (but strictly polynomial) (Fatima, Wooldridge, & Jennings, 2007) and exact (but pseudo-polynomial) algorithms (Mann & Shapley, 1962; Matsui & Matsui, 2000) have been proposed. Chalkiadakis, Elkind, and Wooldridge (2011) provided a comprehensive discussion of this literature.

---

6. Also self-loops are allowed.

Whereas the *choice of representation* has been the foremost consideration for efficient Shapley value computation in the context of conventional coalitional games, in this paper, we face a rather different set of challenges:

- Unlike conventional coalitional games, conciseness is usually not an issue in the networks context. This is because the games that aim to capture network centrality notions are completely specified by (a) the underlying network compactly represented as a graph, and (b) a *concise closed-form characteristic function expression* for evaluating coalition values (see the next section for an example). Rather, the issue here is that the exact specification for the characteristic function is dictated not by computational considerations, but by the real-world application of *game theoretic network centrality*. In other words, the *choice of representation* for Shapley value computation is already fixed by the centrality under consideration.

- Because the games in this paper are designed to reflect network centrality, the characteristic function definition often depends in a highly non-trivial way on the underlying graph structure. Specifically, the value assigned by the characteristic function to each subset of nodes depends not just on the subgraph induced by those nodes, but also on the relationship between that subgraph and the rest of the network. For example, the value assigned to a coalition of nodes may be based on the shortest path lengths to nodes outside the coalition, or it may depend on the relationship between the coalition and its neighbors.

Therefore, the specific challenge we tackle is to *efficiently compute the Shapley value, given a network and a game defined over it, where coalition values for this game are given by a closed-form expression that depends non-trivially on the network*. The key question here is how to take advantage of (a) the network structure, and (b) the functional form for the coalition values, so as to compute Shapley values efficiently, i.e., without the need to enumerate all possible coalitions.

Finally, we conclude this section by mentioning that the Shapley value or other solution concepts from game theory have been applied to other network-related problems. For instance, the application of the Shapley value (and the Nucleolus) to the problem of cost allocation in the electric market transmission system was considered by Zolezzi and Rudnick (2002), though the computational aspects were not discussed. The problem of maximizing the probability of hitting a strategically chosen hidden virtual network by placing a wiretap on a single link of a communication network was analysed by Aziz, Lachish, Paterson, and Savani (2009b). This problem can be viewed as a two-player win-lose (zero-sum) game, a *wiretap game*. The authors not only provide polynomial-time computational results for this game, but also show that one of the (key) strategies is the nucleolus of the simple cooperative spanning connectivity game (Aziz et al., 2009a) mentioned above.

## 3. Preliminaries and Notation

In this section we formally introduce the basic concepts from graph theory and cooperative game theory used throughout the paper. We then look more closely into a sample coalitional game defined over a network and into how the Shapley value of this game can be used as a centrality measure.

A *graph* (or *network*) $G$ consists of *vertices* (or *nodes*) and *edges*, sets of which will be denoted $V(G)$ and $E(G)$, respectively. Every edge from set $E(G)$ connects two vertices in set $V(G)$.[7] By $(u, v)$ we will denote the edge connecting vertices $u, v \in V(G)$. The number of edges incident to a vertex is called a *vertex degree*. The *neighboring vertices* of $v \in V$ are all vertices connected to $v$ in the graph. For a *weighted network* a weight (label) is associated with every edge in $E(G)$. A path is, informally, a sequence of connected edges. The *shortest path* problem is to find a path between two given vertices in which the sum of edge weights is minimized.

We now formalize the notions of coalitional games and the Shapley value. Specifically, let us denote by $A = \{a_1, \ldots, a_{|A|}\}$ the set of players that participate in a coalitional game. A *characteristic function* $\nu :\to \mathbb{R}$ assigns to every coalition $C \subseteq A$ a real number representing the quality of its performance, where it is assumed that $\nu(\emptyset) = 0$. A *characteristic function game* is then a tuple $(A, \nu)$. Assuming that the grand coalition, i.e., the coalition of all the agents in the game, has the highest value and is formed, one of the key questions in coalitional game theory is how to distribute the gain from cooperation among the agents so as to meet certain normative/positive criteria. To this end, Shapley (1953) proposed to evaluate the role of each agent in the game by computing a weighted average of marginal contributions of that agent to all possible coalitions he can belong to. The importance of the Shapley value stems from the fact that it is the unique division scheme that meets four desirable criteria:

(i) *efficiency* — all the wealth available to the agents in the grand coalition is distributed among them;

(ii) *symmetry* — the payoffs to agents do not depend on their identity;

(iii) *null player* — agents with zero marginal contributions to all coalitions receive zero payoff; and

(iv) *additivity* — values of two games sum up to the value computed for the sum of both games.

In order to formalize this concept, let $\pi \in \Pi(A)$ denote a permutation of agents in $A$, and let $C_\pi(i)$ denote the coalition made of all predecessors of agent $a_i$ in $\pi$. More formally, if we denote by $\pi(j)$ the location of $a_j$ in $\pi$, then: $C_\pi(i) = \{a_j \in \pi : \pi(j) < \pi(i)\}$. The Shapley value of $a_i$, denoted $SV_i(\nu)$, is then defined as the average marginal contribution of $a_i$ to coalition $C_\pi(i)$ over all $\pi \in \Pi$ (Shapley, 1953):

$$SV_i(\nu) = \frac{1}{|A|!} \sum_{\pi \in \Pi} [\nu(C_\pi(i) \cup \{a_i\}) - \nu(C_\pi(i))]. \tag{2}$$

---

7. Whereas our main focus in this paper is undirected graphs, we will also show how our results can be readily extended to the case of directed graphs.

Shapley provides the following intuition behind this formula: imagine that the players are to arrive at a meeting point in a random order, and that every player $a_i$ who arrives receives the marginal contribution that his arrival would bring to those already at the meeting point. Now if we average these contributions over all the possible orders of arrival, we obtain $SV_i(\nu)$, $a_i$'s payoff in the game.

The formula in (2) can also be stated in the equivalent, but computationally less involved, form:

$$SV_i(\nu) = \sum_{C \subseteq A \setminus \{a_i\}} \frac{|C|!(|A| - |C| - 1)!}{|A|!} [\nu(C \cup \{a_i\}) - \nu(C)]. \qquad (3)$$

In our network context, we will use $G$ to define a coalitional game $(V(G), \nu)$ with set of agents $A = V(G)$ and characteristic function $\nu$. Here the agents of the coalitional game are the vertices of the graph $G$. Thus, a coalition of agents $C$ is simply any subset of $V(G)$. Furthermore, the characteristic function $\nu : 2^{V(G)} \to \mathbb{R}$ can be any function that depends on the graph $G$ as long as it satisfies the condition $\nu(\emptyset) = 0$. We use the phrase "value of coalition $C$" to informally refer to $\nu(C)$.

We will first consider a sample characteristic function game defined over a network, as well as its Shapley value that becomes a centrality measure. We will then discuss the advantages of the game theoretic network centrality over conventional measures.

In more detail, consider the notion of "closeness centrality" of a node in a graph $G(V, E)$, which is traditionally defined as the reciprocal of the average distance of that node from other (reachable) nodes in the graph (Koschützki et al., 2005). This definition captures the intuitive idea that a node "in close proximity to many other nodes" is more valuable by virtue of its central location, and hence should be assigned a higher centrality score.

The above measure, however, fails to recognize the importance of combinations of nodes. For example, consider a typical application of closeness centrality: that of disseminating a piece of information to all nodes in the network. At any time point $t$ in the dissemination process, define the random variable $C_t$ to be the subset of nodes actively involved in propagating the information. In this situation, a new node added to $C_t$ would make maximum contribution to the diffusion of information only if it is "in close proximity to nodes that are not currently in close proximity to any node in $C_t$". Thus, while conventional closeness centrality only takes into account *average proximity to all other nodes*, the actual importance of a node in actual applications is based on a very different measure: *proximity to nodes that are not in close proximity to the random variable $C_t$*.

We now show how coalitional game theory can be used to construct a centrality measure that more faithfully models the above situation. Let $C$ be an arbitrary subset of nodes from the given network $G(V, E)$. Then, for every such $C$, assign a value $\nu(C)$ given by

$$\nu(C) = \sum_{v \in V(G)} \frac{1}{1 + \min\{d(u, v) | u \in C\}},$$

where $d(u, v)$ is the distance between nodes $u$ and $v$ (measured as the shortest path length between $u$ and $v$ in graph $G$).

The map $\nu$ defined above captures a fundamental centrality notion: that the *intrinsic value* of a subset of nodes $C$ in the context of such an application as information dissemination is proportional to the overall proximity of the nodes in $C$ to the other nodes in the network. In effect, the map $\nu$ carries the original definition of closeness centrality to a global level, where a measure of importance is assigned to every possible combination of nodes.

The map $\nu$ above is therefore a *characteristic function* for a coalitional game, where each vertex of the network is viewed as an agent playing the game. It follows that if a node $v$ has a high Shapley value in this game, it is likely that $v$ would "contribute more" to an arbitrary randomly chosen coalition of nodes $C$ in terms of increasing the proximity of $C$ to other nodes on the network. Thus, computing the Shapley values of this game yields a centrality score for each vertex that is a much-improved characterization of closeness centrality.

The only difficulty in adopting such a game-theoretically inspired centrality measure is the previously mentioned problem of *exponential complexity in the number of agents*. In the next section, we show how to overcome this difficulty and compute the Shapley value for many centrality applications (including the above formulation) in time polynomial in the size of the network.

## 4. Algorithms for Shapley Value-Based Network Centrality

In this section, we present five characteristic function formulations $\nu(C)$, each designed to convey a specific centrality notion. As already mentioned in the introduction, a common element of all these formulations is that they aim to quantify, albeit in a different way, the *sphere of influence* of the coalition $C$ over the other nodes in the network. Specifically, in our first game formulation, we start with the simplest possible idea that the sphere of influence of a coalition of nodes $C$ is the set of all nodes immediately reachable (within one hop) from $C$. Subsequent games further generalize this notion. In particular, the second formulation specifies a more sophisticated sphere of influence: one that includes only those nodes which are immediately reachable in at least $k$ different ways from $C$. The other three formulations extend the notion of sphere of influence to weighted graphs. The third game defines sphere of influence as the set of all nodes within a cutoff distance of $C$ (as measured by shortest path lengths on the weighted graph). The fourth formulation is an extreme generalization: it allows the sphere of influence of $C$ to be specified by an arbitrary function $f(.)$ of the distance between $C$ and the other nodes. The final formulation is a straightforward extension of the second game, to the case of weighted networks.

The relationships among all five games are graphically presented in Figure 4.

### 4.1 Game 1: $\nu_1(C) = \#$agents at most 1 degree away

Let $G(V,E)$ be an unweighted, undirected network. We first define the "fringe" of a subset $C \subseteq V(G)$ as the set $\{v \in V(G) : v \in C$ (or) $\exists u \in C$ such that $(u,v) \in E(G)\}$, i.e., the fringe of a coalition includes all nodes reachable from the coalition in at most one hop.

Based on the fringe, we define the coalitional game $g_1(V(G), \nu_1)$ with respect to the network $G(V,E)$ by the characteristic function $\nu_1 : 2^{V(G)} \to \mathbb{R}$ given by

$$\nu_1(C) = \begin{cases} 0 & \text{if } C = \emptyset \\ \text{size}(\text{fringe}(C)) & \text{otherwise.} \end{cases}$$

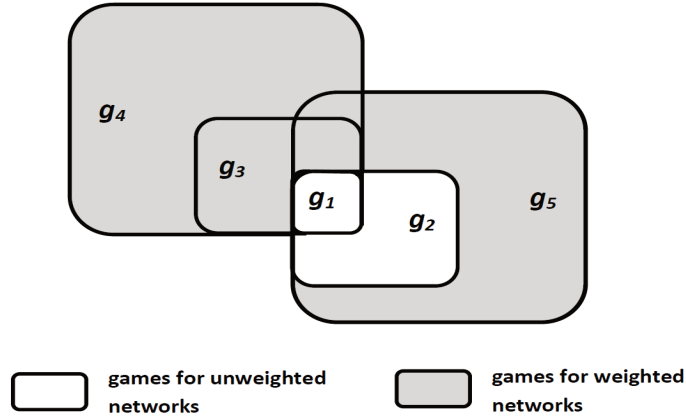

Figure 4: Euler diagram showing the relationships among all five games considered in this paper. Specifically, game $g_2$ generalizes $g_1$; $g_3$ generalizes $g_1$ and is further generalized by $g_4$; $g_5$ generalizes $g_2$. Finally, we note that there are certain instances of games that can be represented as $g_3$, $g_4$ and $g_5$.

.

The above game was applied by Suri and Narahari (2008) to find out influential nodes in social networks and it was shown to deliver very promising results concerning the target set selection problem (see Kempe, Kleinberg, & Tardos, 2003). It is therefore desired to compute the Shapley values of all nodes for this game. We shall now present an exact formula for this computation rather than obtaining it through Monte Carlo simulation as was done by Suri and Narahari.

In more detail, to evaluate the Shapley value of node $v_i$, consider all possible permutations of the nodes in which $v_i$ would make a positive marginal contribution to the coalition of nodes occurring before itself. Let the set of nodes occurring before node $v_i$ in a random permutation of nodes be denoted $C_i$. Let the neighbours of node $v_i$ in the graph $G(V, E)$ be denoted $N_G(v_i)$ and the degree of node $v_i$ be denoted $deg_G(v_i)$.

The key question to ask is: what is the necessary and sufficient condition for node $v_i$ to "marginally contribute node $v_j \in N_G(v_i) \cup \{v_i\}$ to fringe$(C_i)$"? Clearly, this happens if and only if neither $v_j$ nor any of *its* neighbours are present in $C_i$. Formally, $(N_G(v_j) \cup \{v_j\}) \cap C_i = \emptyset$.

Now we are going to show that the above condition holds with probability $\frac{1}{1+deg_G(v_j)}$.

**Proposition 1.** *The probability that in a random permutation none of the vertices from $N_G(v_j) \cup \{v_j\}$ occurs before $v_i$, where $v_i$ and $v_j$ are neighbours, is $\frac{1}{1+deg_G(v_j)}$.*

---

**Algorithm 1:** Computing the Shapley value for Game 1

---

**Input**: Unweighted graph $G(V, E)$
**Output**: Shapley values of all nodes in $V(G)$ for game $g_1$

**foreach** $v \in V(G)$ **do**
    $\text{SV}[v] = \frac{1}{1 + deg_G(v)}$;
    **foreach** $u \in N_G(v)$ **do**
        $\text{SV}[v] \mathrel{+}= \frac{1}{1 + deg_G(u)}$;
    **end**
**end**
return SV;

---

*Proof.* We need to count the number of permutations that satisfy:

$$\forall_{v \in (N_G(v_j) \cup \{v_j\})} \pi(v_i) < \pi(v). \tag{4}$$

To this end:

- Let us choose $|(N_G(v_j) \cup \{v_j\}|$ positions in the sequence of all elements from $V$. We can do this in $\binom{|V|}{1 + deg_G(v_j)}$ ways.

- Then, in the last $deg_G(v_j)$ chosen positions, place all elements from $(N_G(v_j) \cup \{v_j\}) \setminus \{v_i\}$. Directly before these, place the element $v_i$. The number of such line-ups is $(deg_G(v_j))!$.

- The remaining elements can be arrange in $(|V| - (1 + deg_G(v_j)))!$ different ways.

All in all, the number of permutations satisfying condition (4) is:

$$\binom{|V|}{1 + deg_G(v_j)}(deg_G(v_j))!(|V| - (1 + deg_G(v_j)))! = \frac{|V|!}{1 + deg_G(v_j)};$$

thus, the probability that one of such permutations is randomly chosen is $\frac{1}{1 + deg_G(v_j)}$. $\qquad\square$

Now, denote by $B_{v_i, v_j}$ the Bernoulli random variable that $v_i$ marginally contributes $v_j$ to fringe($C_i$). From the above, we have:

$$E[B_{v_i, v_j}] = \Pr[(N_G(v_j) \cup \{v_j\}) \cap C_i = \emptyset] = \frac{1}{1 + deg_G(v_j)}.$$

Therefore, $SV_{g_1}(v_i)$, which is the expected marginal contribution of $v_i$, is given by:

$$SV_{g_1}(v_i) = \sum_{v_j \in \{v_i\} \cup N_G(v_i)} E[B_{v_i, v_j}] = \sum_{v_j \in \{v_i\} \cup N_G(v_i)} \frac{1}{1 + deg_G(v_j)}, \tag{5}$$

which is an *exact closed-form expression* for computing the Shapley value of each node on the network.

It is possible to derive some intuition from the above formula. If a node has a high degree, the number of terms in its Shapley value summation above is also high. But the terms themselves will be inversely related to the degree of neighboring nodes. This gives the intuition that a node will have high centrality not only when its degree is high, but also whenever its degree tends to be higher in comparison to the degree of its neighboring nodes. In other words, *power comes from being connected to those who are powerless*, a fact that is well-recognized by the centrality literature (e.g., Bonacich, 1987). Following the same reasoning, we can also easily predict how dynamic changes to the network, such as adding or removing an edge, would influence the Shapley value.[8] Adding an edge between a powerful and a powerless node will add even more power to the former and will decrease the power of the latter. Naturally, removing an edge would have the reverse effect.

Interestingly, although game $g_1$ is quite different from the induced-subgraph representation of Deng and Papadimitriou (1994), the formula for $SV_{g_1}(v_i)$ is, to some extent, similar to formula (1). In particular, in both cases, the Shapley value of a node depends solely on the set of its immediate neighbours. Moreover, in both cases, it is a linear combination of fractions involving in the numerator the weight of edges between the node and its neighbours.[9] The difference is in the denominators, where in our case it depends on the degree of the involved nodes. We will see that the next two games considered in this paper yield comparable to $g_1$ closed-form expressions for the Shapley value.

Algorithm 1 directly implements expression (5) to compute the exact Shapley values of all nodes in the network. It cycles through all nodes and their neighbours, so its running time is $O(|V| + |E|)$.

Finally, we note that Algorithm 1 can be adopted to directed graphs with a couple of simple modifications. Specifically, in order to capture how many nodes we can access a given node from, the degree of a node should be replaced with indegree. Furthermore, a set of neighbours of a given node $v$ should consist of those nodes to which an edge is directed from $v$.

## 4.2 Game 2: $\nu_2(C) = \#$**agents with at least $k$ neighbors in $C$**

We now consider a more general game formulation for an unweighted graph $G(V, E)$, where the value of a coalition includes the number of agents that are either in the coalition or are adjacent to at least $k$ agents who are in the coalition. Formally, we consider game $g_2$ characterised by $\nu_2 : 2^{V(G)} \to \mathbb{R}$, where

$$\nu_2(C) = \begin{cases} 0 & \text{if } C = \emptyset \\ |\{v : v \in C \text{ (or) } |N_G(v) \cap C| \geq k\}| & \text{otherwise.} \end{cases}$$

---

8. Many real-life networks are in fact dynamic and the challenge of developing fast streaming algorithms has recently attracted considerable attention in the literature (Lee, Lee, Park, Choi, & Chung, 2012).

9. Note that in case of $g_1$ the weight of any edge is 1.

The second game is an instance of the General Threshold Model that has been widely studied in the literature (e.g., Kempe et al., 2005; Granovetter, 1978). Intuitively, in this model each node can become active if a monotone activation function reaches some threshold. The instance of this problem has been proposed by Goyal, Bonchi, and Lakshmanan (2010), where the authors introduced a method of learning influence probabilities in social networks (from users' action logs). However, in many realistic situations much less information is available about a network so it is not possible to assess specific probabilities with which individual nodes become active. Consequently, much simpler models are studied. Bikhchandani et al. (1992), for instance, "consider a teenager deciding whether or not to try drugs. A strong motivation for trying out drugs is the fact that friends are doing so. Conversely, seeing friends reject drugs could help persuade the teenager to stay clean". This situation is modelled by the second game; the threshold for each node is $k$ and the activation function is $f(S) = |S|$. Another example is viral marketing or innovation diffusion analysis. Again, in this application, it is often assumed that an agent will "be influenced" only if at least $k$ of his neighbors have already been convinced (Valente, 1996). Note that this game reduces to game $g_1$ for $k = 1$.

Adopting notation from the previous subsection, we again ask: what is the necessary and sufficient condition for node $v_i$ to marginally contribute node $v_j \in N_G(v_i) \cup \{v_i\}$ to the value of the coalition $C_i$?

Clearly, if $deg_G(v_j) < k$, we have $E[B_{v_i,v_j}] = 1$ for $v_i = v_j$ and 0 otherwise. For $deg_G(n_j) \geq k$, we split the argument into two cases. If $v_j \neq v_i$, the condition for marginal contribution is that exactly $(k-1)$ neighbors of $v_j$ already belong to $C_i$ and $v_j \notin C_i$. On the other hand, if $v_j = v_i$, the marginal contribution occurs if and only if $C_i$ originally consisted of at most $(k-1)$ neighbors of $v_j$. So for $deg_G(v_j) \geq k$ and $v_j \neq v_i$, we need to determine the appropriate probability.

**Proposition 2.** *The probability that in a random permutation exactly $k-1$ neighbours of $v_j$ occur before $v_i$, and $v_j$ occurs after $v_i$, is: $\frac{1 + deg_G(v_j) - k}{deg_G(v_j)(1 + deg_G(v_j))}$, where $v_j$ and $v_i$ are neighbors and $deg_G(v_j) \geq k$.*

*Proof.* We need to count the number of permutations that satisfy:

$$\exists!_{K \subseteq N_G(v_j)} \Big\{ |K| = k - 1 \ \wedge \ \forall_{v \in K} \big\{ \pi(v) < \pi(v_i) \big\} \ \wedge$$
$$\forall_{v \in N_G(v_j) \setminus K} \big\{ \pi(v_i) \leq \pi(v) \big\} \ \wedge \ \pi(v_i) < \pi(v_j) \Big\}. \tag{6}$$

To this end:

- Let us choose $|(N_G(v_j) \cup \{v_j\}|$ positions in the sequence of all elements from $V$. We can do this in $\binom{|V|}{1 + deg_G(v_j)}$ ways.

- Then, choose $k - 1$ elements from the set $(N_G(v_j) \setminus \{v_i\}$. The number of such choices is $\binom{deg_G(v_j) - 1}{k - 1}$.

---

**Algorithm 2:** Computing the Shapley value for Game 2

---

**Input**: Unweighted graph $G(V, E)$, positive integer $k$
**Output**: Shapley value of all nodes in $V(G)$ for game $g_2$

**foreach** $v \in V(G)$ **do**
$\quad$ SV[$v$] = min(1, $\frac{k}{1+deg_G(v)}$);
$\quad$ **foreach** $u \in N_G(v)$ **do**
$\quad\quad$ SV[$v$] += max(0, $\frac{deg_G(u)-k+1}{deg_G(u)(1+deg_G(u))}$);
$\quad$ **end**
**end**
return SV;

---

- Then, in the first $k - 1$ chosen positions, place all elements chosen in previous step. Directly after those, place the element $v_i$, and then the remaining vertices chosen in the first step. The number of such line-ups is $(k-1)!(1 + deg_G(v_j) - k)!$.

- The remaining elements can be arrange in $(|V| - deg_G(v_j) - 1)!$ different ways.

Taking all the above together, the number of permutations satisfying condition (6) is:

$\binom{|V|}{1+deg_G(v_j)}\binom{deg_G-1}{k-1}(k-1)!(1 + deg_G(v_j) - k)!(|V| - deg_G(v_j) - 1)! = \frac{|V|!(1+deg_G(v_j)-k)}{deg_G(v_j)(1+deg_G(v_j))}$;

thus, the probability that one of such permutations is randomly chosen is $\frac{1+deg_G(v_j)-k}{deg_G(v_j)(1+deg_G(v_j))}$.
$\qquad\qquad\qquad\qquad\qquad\qquad\qquad\qquad\qquad\qquad\qquad\qquad\qquad\qquad\qquad\qquad\qquad\qquad$ $\square$

Using Proposition 2 we obtain:

$$E[B_{v_i, v_j}] = \frac{1 + deg_G(v_j) - k}{deg_G(v_j)(1 + deg_G(v_j))}.$$

And for $deg_G(v_i) \geq k$ and $v_j = v_i$, we have:

$$E[B_{v_i, v_i}] = \sum_{n=0}^{k-1} \frac{1}{1 + deg_G(v_i)} = \frac{k}{1 + deg_G(v_i)}.$$

As before, the Shapley values are given by substituting the above formulae into:

$$SV_{g_2}(v_i) \;=\; \sum_{v_j \in N_G(v_i) \cup \{v_i\}} E[B_{v_i, v_j}].$$

Although this game is a generalization of game $g_1$, it can still be solved to obtain the Shapley values of all nodes in $O(|V| + |E|)$ time, as formalised by Algorithm 2.

624

An even more general formulation of the game is possible by allowing $k$ to be a function of the agent, i.e., each node $v_i \in V(G)$ is assigned its own unique attribute $k(v_i)$. This translates to an application of the form: agent $i$ is convinced if and only if at least $k_i$ of his neighbors are convinced, which is a frequently used model in the literature (Valente, 1996).

The above argument does not use the fact that $k$ is constant across all nodes. So this generalized formulation can be solved by a simple modification to the original Shapley value expression:

$$SV(v_i) = \frac{k(v_i)}{1 + deg_G(v_i)} + \sum_{v_j \in N_G(v_i)} \frac{1 + deg_G(v_j) - k(v_j)}{deg_G(v_j)(1 + deg_G(v_j))}.$$

The above equation (which is also implementable in $O(|V| + |E|)$ time) assumes that $k(v_i) \leq 1 + deg_G(v_i)$ for all nodes $v_i$. This condition can be assumed without loss of generality because all cases can still be modelled (we set $k(v_i) = 1 + deg_G(v_i)$ for the extreme case where node $v_i$ is never convinced no matter how many of its neighbors are already convinced).

Finally, we note that Algorithm 2 can be adapted to a case of directed graphs along the same lines as Algorithm 1.

### 4.3 Game 3: $\nu_3(C) = \#$**agents at most $d_{cutoff}$ away**

Hitherto, our games have been confined to unweighted networks. But in many applications, it is necessary to model real-world networks as weighted graphs. For example, in the co-authorship network mentioned in the introduction, each edge is often assigned a weight proportional to the number of joint publications the corresponding authors have produced (Newman, 2001).

This subsection extends game $g_1$ to the case of weighted networks. Whereas game $g_1$ equates $\nu(C)$ to the number of nodes located within one hop of some node in $C$, our formulation in this subsection equates $\nu(C)$ to the number of nodes located within a distance $d_{cutoff}$ of some node in $C$. Here, distance between two nodes is measured as the length of the shortest path between the nodes in the given weighted graph $G(V, E, W)$, where $W : E \to \mathbb{R}^+$ is the weight function.

Formally, we define game $g_3$, where for each coalition $C \subseteq V(G)$,

$$\nu_3(C) = \begin{cases} 0 & \text{if } C = \emptyset \\ \text{size}(\{v_i : \exists v_j \in C \mid \text{distance}(v_i, v_j) \leq d_{cutoff}\}) & \text{otherwise.} \end{cases}$$

Clearly, $g_3$ can be used in all the settings where $g_1$ is applicable; for instance, in the diffusion of information in social networks or to analyse research collaboration networks (e.g., Suri & Narahari, 2010, 2008). Moreover, as a more general game, $g_3$ provides additional modelling opportunities. For instance, Suri and Narahari (2010) suggest that a "*more intelligent*" way for sieving nodes in the neighbourhood would improve their algorithm for solving the target selection problem (*top-k* problem). Now, $g_3$ allows us to define a different *cutoff* distance

---

**Algorithm 3:** Computing the Shapley value for Game 3

---

**Input**: Weighted graph $G(V, E, W)$, $d_{cutoff} > 0$
**Output**: Shapley value of all nodes in $G$ for game $g_3$

**foreach** $v \in V(G)$ **do**
    DistanceVector D = Dijkstra$(v, G)$;
    extNeighbors$(v) = \emptyset$; extDegree$(v) = 0$;
    **foreach** $u \in V(G)$ *such that* $u \neq v$ **do**
        **if** $D(u) \leq d_{cutoff}$ **then**
            extNeighbors$(v)$.push$(u)$;
            extDegree$(v)$++;
        **end**
    **end**
**end**
**foreach** $v \in V(G)$ **do**
    SV$[v] = \frac{1}{1 + extDegree(v)}$;
    **foreach** $u \in extNeighbors(v)$ **do**
        SV$[v]$ += $\frac{1}{1 + extDegree(u)}$;
    **end**
**end**
return SV;

---

for each node in Suri and Narahari's setting. Furthermore, $g_3$ is a specific case of the more general model $g_4$ which will be discussed in next subsection.

We shall now show that even this highly general centrality game $g_3$ is amenable to analysis which yields an exact formula for the Shapley value. However, in this case the algorithm for implementing the formula is not linear in the size of the network, but has $O(|V||E| + |V|^2 \log|V|)$ complexity.

Before deriving the exact Shapley value formula, we introduce some extra notation. Define the *extended neighborhood* $N_G(v_j, d_{cutoff}) = \{v_k \neq v_j : \text{distance}(v_k, v_j) \leq d_{cutoff}\}$, i.e., the set of all nodes whose distance from $v_j$ is at most $d_{cutoff}$. Denote the size of $N_G(v_j, d_{cutoff})$ by $deg_G(v_j, d_{cutoff})$. With this notation, the necessary and sufficient condition for node $v_i$ to marginally contribute node $v_j$ to the value of coalition $C_i$ is: distance$(v_i, v_j) \leq d_{cutoff}$ and distance$(v_j, v_k) > d_{cutoff}$ $\forall v_k \in C_i$. That is, neither $v_j$ nor any node in *its* extended neighborhood should be present in $C_i$. From the discussion in previous subsections and Proposition 1, we know that the probability of this event is exactly $\frac{1}{1 + deg_G(v_j, d_{cutoff})}$. Therefore, the exact formula for the Shapley value of node $v_i$ in game $g_3$ is:

$$SV_{g_3}(v_i) \quad = \sum_{v_j \in \{v_i\} \cup N_G(v_i, d_{cutoff})} \frac{1}{1 + deg_G(v_j, d_{cutoff})}.$$

Algorithm 3 works as follows: for each node $v$ in the network $G(V, E)$, the extended neighborhood $N_G(v, d_{cutoff})$ and its size $deg_G(v, d_{cutoff})$ are first computed using Dijkstra's algorithm in $O(|E| + |V| \log |V|)$ time (Cormen, 2001). The results are then used to directly implement the above equation, which takes maximum time $O(|V|^2)$. In practice this step runs much faster because the worst case situation only occurs when every node is reachable from every other node within $d_{cutoff}$. Overall the complexity is $O(|V||E| + |V|^2 \log |V|)$.

Furthermore, to deal with directed graphs we need to redefine the notion of *extDegree* and *extNeighbors* for a given node $u$ in Algorithm 3. The former will be the number of vertices from which the distance to $u$ is smaller than, or equal to, $d_{cutoff}$. The latter will be the set of nodes whose distance from $u$ is at most $d_{cutoff}$.

Finally, we make the observation that the above proof does not depend on $d_{cutoff}$ being constant across all nodes. Indeed, each node $v_i \in V(G)$ may be assigned its own unique value $d_{cutoff}(v_i)$, where $\nu(C)$ would be the number of agents $v_i$ who are within a distance $d_{cutoff}(v_i)$ from $C$. For this case, the above proof gives:

$$SV(v_i) \;=\; \sum_{\substack{v_j : \text{distance}(v_i, v_j) \\ \le d_{cutoff}(v_j)}} \frac{1}{1 + deg_G(v_j, d_{cutoff}(v_j))}.$$

### 4.4 Game 4: $\nu_4(C) = \sum_{v_i \in V(G)} f(\text{distance}(v_i, C))$

This subsection further generalizes game $g_3$, again taking motivation from real-life network problems. In game $g_3$, all agents at distances $d_{\text{agent}} \le d_{cutoff}$ contributed equally to the value of a coalition. However, this assumption may not always hold true because in some applications we intuitively expect agents closer to a coalition to contribute more to its value. For instance, we expect a Facebook user to exert more influence over his immediate circle of friends than over "friends of friends", even though both may satisfy the $d_{cutoff}$ criterion. Similarly, we expect a virus-affected computer to infect a neighboring computer more quickly than a computer two hops away.

In general, we expect that an agent at distance $d$ from a coalition would contribute $f(d)$ to its value, where $f(.)$ is a positive valued decreasing function of its argument. More formally, we define game $g_4$, where the value of a coalition $C$ is given by:

$$\nu_4(C) = \begin{cases} 0 & \text{if } C = \emptyset \\ \sum_{v_i \in V(G)} f(d(v_i, C)) & \text{otherwise,} \end{cases}$$

where $d(v_i, C)$ is the minimum distance: $\min\{\text{distance}(v_i, v_j) | v_j \in C\}$.

We note that it is possible to solve for the Shapley value in the above formulation by constructing an MC-Nets representation (see Section 2 for more details on this formalism). Indeed, the combinatorial structure of networks is to a certain extent similar to the structure of MC-Nets. Consequently, the existence of a polynomial algorithm to compute the Shapley value for MC-Nets strongly suggests that polynomial algorithms could be developed for games defined over networks. Our results in this paper demonstrate that this is indeed the case. However, it should be underlined that our approach to compute the Shapley value is different from that applied in MC-Nets. This is because in our solutions we focus on computing the expected contribution of every node in a random permutation of nodes and not on disaggregating the game into a collection of "simple", easily solvable, games as it is done in MC-Nets. The difference in both approaches is clearly visible in the case of $g_3$. Here, the MC-Nets would have $O(|V|^3)$ rules, whereas in the discussion below, we propose a more efficient algorithm for $g_3$ that runs in $O(|V||E| + |V|^2 \log |V|)$. This is a considerable improvement because most real-world networks are sparse, i.e., $E \sim O(|V|)$ (Reka & Barabási, 2002).

In the case of $g_3$, the key question to ask is: what is the expected value of the marginal contribution of $v_i$ through node $v_j \neq v_i$ to the value of coalition $C_i$? Let this marginal contribution be denoted $MC(v_i, v_j)$. Clearly:

$$MC(v_i, v_j) = \begin{cases} 0 & \text{if distance}(v_i, v_j) \geq d(v_j, C_i) \\ f(\text{distance}(v_i, v_j)) - f(d(v_j, C_i)) & \text{otherwise.} \end{cases}$$

Let $D_{v_j} = \{d_1, d_2 ... d_{|V|-1}\}$ be the distances of node $v_j$ from all other nodes in the network, sorted in increasing order. Let the nodes corresponding to these distances be $\{w_1, w_2 ... w_{|V|-1}\}$, respectively. Let $k_{ij} + 1$ be the number of nodes (out of these $|V| - 1$) whose distances to $v_j$ are $\leq \text{distance}(v_i, v_j)$. Let $w_{k_{ij}+1} = v_i$ (i.e., among all nodes that have the same distance from $v_j$ as $v_i$, $v_i$ is placed last in the increasing order).

We use *literal* $w_i$ to mean $w_i \in C_i$ and the literal $\overline{w_i}$ to mean $w_i \notin C_i$. Define a sequence of boolean variables $p_k = \overline{v_j} \wedge \overline{w_1} \wedge \overline{w_2} \wedge ... \wedge \overline{w_k}$ for each $0 \leq k \leq |V| - 1$. Finally denote expressions of the form $MC(v_i, v_j | F)$ to mean the marginal contribution of $v_i$ to $C_i$ through $v_j$ given that the coalition $C_i$ satisfies the boolean expression $F$.

$$MC(v_i, v_j | p_{k_{ij}+1} \wedge w_{k_{ij}+2}) = f(d_{k_{ij}+1}) - f(d_{k_{ij}+2}),$$
$$MC(v_i, v_j | p_{k_{ij}+2} \wedge w_{k_{ij}+3}) = f(d_{k_{ij}+1}) - f(d_{k_{ij}+3}),$$
$$\vdots \qquad \vdots \qquad \vdots$$
$$MC(v_i, v_j | p_{|V|-2} \wedge w_{|V|-1}) = f(d_{k_{ij}+1}) - f(d_{|V|-1}),$$
$$MC(v_i, v_j | p_{|V|-1}) = f(d_{k_{ij}+1}).$$

With this notation, we obtain expressions for $MC(v_i, v_j)$ by splitting over the above *mutually exclusive* and *exhaustive* (i.e., covering all possible non-zero marginal contributions) cases.

Now, we need to determine the probability of $\Pr(p_k \wedge w_{k+1})$.

**Proposition 3.** *The probability that in a random permutation none of the nodes from $\{v_j, w_1, \ldots, w_k\}$ occur before $v_i$ and the node $w_{k+1}$ occurs before $v_i$ is $\frac{1}{(k+1)(k+2)}$.*

*Proof.* Let us count the number of permutations that satisfy:

$$\forall_{v \in \{v_j, w_1, ..., w_k\}} \pi(v_i) < \pi(v) \ \wedge \ \pi(v_i < \pi(w_{k+1}). \tag{7}$$

To this end:

- Let us choose $|\{v_j, w_1, \ldots, w_k\} \cup \{v_j\} \cup \{w_{k+1}\}|$ positions in the sequence of all elements from $V$. We can do this in $\binom{|V|}{k+3}$ ways.

- Then, in the last $k+1$ chosen positions, we place all elements from $\{v_j, w_1, \ldots, w_k\}$. Directly before these, we place the element $v_i$, and then vertex $w_{k+1}$. The number of such line-ups is $(k+1)!$.

- The remaining elements can be arrange in $(|V| - (k+3))!$ different ways.

Thus, the number of permutations satisfying (7) is:

$$\binom{|V|}{k+3}(k+1)!(|V| - (k+3))! = \frac{|V|!}{(k+1)(k+2)},$$

and the probability that one of such permutations is randomly chosen is $\frac{1}{(k+1)(k+2)}$. $\qquad \square$

With the above proposition we find that:

$$\Pr(p_k \wedge w_{k+1}) \frac{1}{(k+1)(k+2)} \ \forall \ 1 + k_{ij} \leq k \leq |V| - 2.$$

Using the $MC(v_i, v_j)$ equations and the probabilities $\Pr(p_k \wedge w_{k+1})$:

$$\begin{aligned} E[MC(v_i, v_j)] &= \left[ \sum_{k=1+k_{ij}}^{|V|-2} \frac{f(\text{distance}(v_i, v_j)) - f(d_{k+1})}{(k+1)(k+2)} \right] + \frac{f(\text{distance}(v_i, v_j))}{|V|} \\ &= \frac{f(\text{distance}(v_i, v_j))}{k_{ij} + 2} - \sum_{k=k_{ij}+1}^{|V|-2} \frac{f(d_{k+1})}{(k+1)(k+2)}. \end{aligned}$$

For $v_i = v_j$, a similar analysis produces:

$$E[MC(v_i, v_i)] = f(0) - \sum_{k=0}^{|V|-2} \frac{f(d_{k+1})}{(k+1)(k+2)}.$$

Finally the exact Shapley value is given by:

$$SV_{g_4}(v_i) = \sum_{v_j \in V(G)} E[MC(v_i, v_j)].$$

Algorithm 4 implements the above formulae. For each vertex $v$, a vector of distances to every other vertex is first computed using Dijkstra's algorithm (Cormen, 2001). This yields a vector $D_v$ that is already sorted in increasing order. This vector is then traversed in

---

**Algorithm 4:** Computing the Shapley value for Game 4

---

**Input**: Weighted graph $G(V, E, W)$, function $f : \mathbb{R}^+ \to \mathbb{R}^+$
**Output**: Shapley value of all nodes in $G$ for game $g_4$

**Initialise:** $\forall v \in V(G)$ **set** SV$[v] = 0$;
**foreach** $v \in V(G)$ **do**
    [Distances D, Nodes w] = Dijkstra$(v, G)$;
    sum = 0; index = |V|-1; prevDistance = -1, prevSV = -1;
    **while** *index > 0* **do**
        **if** *D(index) == prevDistance* **then**
            currSV = prevSV;
        **else**
            currSV = $\frac{f(D(index))}{1+index} - sum$;
        **end**
        SV[w(index)] += currSV;
        sum += $\frac{f(D(index))}{index(1+index)}$;
        prevDistance = D(index), prevSV = currSV;
        index--;
    **end**
    SV$[v]$ += f(0) $-$ sum;
**end**
return SV;

---

reverse, to compute the backwards cumulative sum $\sum \frac{f(d_{k+1})}{(k+1)(k+2)}$. At each step of the backward traversal, the Shapley value of the appropriate node $w$ is updated according to the $E[MC(w, v)]$ equation. After the traversal, the Shapley value of $v$ itself is updated according to the $E[MC(v, v)]$ equation. This process is repeated for all nodes $v$ so that at the end of the algorithm, the Shapley value is computed exactly in $O(|V||E| + |V|^2 \log |V|)$ time.

Our final observation is that Algorithm 4 works also for directed graphs as long as we use the appropriate version of Dijkstra's algorithm (see, e.g., Cormen, 2001).

### 4.5 Game 5: $\nu_5(C) = \#$**agents with** $\sum$(**weights inside** $C$) $\geq W_{cutoff}$(**agent**)

In this subsection, we generalize game $g_2$ for the case of weighted networks. Given a positive weighted network $G(V, E, W)$ and a value $W_{cutoff}(v_i)$ for every node $v_i \in V(G)$, we first define $W(v_j, C) = \sum_{v_i \in C} W(v_j, v_i)$ for every coalition $C$, where $W(v_i, v_j)$ is the weight of the edge between nodes $v_i$ and $v_j$ (or 0 if there is no such edge). With this notation, we define game $g_5$ by the characteristic function:

$$
\nu_5(C) = \begin{cases} 0 & \text{if } C = \emptyset \\ \text{size}(\{v_i : v_i \in C \ (or) \ W(v_i, C) \geq W_{cutoff}(v_i)\}) & \text{otherwise.} \end{cases}
$$

---

**Algorithm 5:** Computing the Shapley value for Game 5

---

**Input**: Weighted network $G(V, E, W)$, cutoffs $W_{cutoff}(v_i)$ for each $v_i \in V(G)$
**Output**: Shapley value of all nodes in $G$ for game $g_5$

**foreach** $v_i \in V(G)$ **do**
    compute and store $\alpha_i$ and $\beta_i$;
**end**
**foreach** $v_i \in V(G)$ **do**
    SV$[v_i] = 0$;
    **foreach** $m$ *in* $0$ *to* $deg_G(v_i)$ **do**
        compute $\mu = \mu(X_m^{ii})$, $\sigma = \sigma(X_m^{ii})$, $p = \Pr\{\mathcal{N}(\mu, \sigma^2) < W_{cutoff}(v_i)\}$;
        SV$[v_i] \mathrel{+}= \frac{p}{1+deg_G(v_i)}$;
    **end**
    **foreach** $v_j \in N_G(v_i)$ **do**
        p = 0;
        **foreach** $m$ *in* $0$ *to* $deg_G(v_j) - 1$ **do**
            compute $\mu = \mu(X_m^{ij})$, $\sigma = \sigma(X_m^{ij})$ and $z = Z_m^{ij}$;
            $p \mathrel{+}= z \frac{deg_G(v_j) - m}{deg_G(v_j)(deg_G(v_j)+1)}$;
        **end**
        SV$[v_i] \mathrel{+}=$ p;
    **end**
**end**
return SV;

---

The formulation above has applications in, for instance, the analysis of information diffusion, adoption of innovations, and viral marketing. Indeed, many cascade models of such phenomena on weighted graphs have been proposed (e.g., Granovetter, 1978; Kempe et al., 2003; Young, 2006) which work by assuming that an agent will change state from "inactive" to "active" if and only if the sum of the weights to all active neighbors is at least equal to an agent-specific cutoff.

Although we have not been able to come up with an exact formula for the Shapley value in this game[10], our analysis yields an approximate formula which was found to be accurate in practice.

In more detail, we observe that node $v_i$ marginally contributes node $v_j \in N_G(v_i)$ to the value of coalition $C_i$ if and only if $v_j \notin C_i$ and $W_{cutoff}(v_j) - W(v_i, v_j) \le W(v_j, C_i) < W_{cutoff}(v_j)$. Let us denote by $B_{v_i, v_j}$ the Bernoulli random variable corresponding to this event. We will need the following additional notation:

- let $N_G(v_j) = \{v_i, w_1, w_2 ... w_{deg_G(v_j) - 1}\}$;

---

10. Computing the Shapley value for this game involves determining whether the sum of weights on specific edges, adjacent to a random coalition, exceeds the threshold. This problem seems to be at least as hard as computing the Shapley value in weighted voting games, which is #P-Complete (Elkind et al., 2009).

- let the weights of edges between $v_j$ and each of the nodes in $N_G(v_j)$ be $W_j = \{W(v_i, v_j), W_1, W_2...W_{deg_G(v_j)-1}\}$ in that order;

- let $\alpha_j$ be the sum of all the weights in $W_j$ and $\beta_j$ be the sum of the squares of all the weights in $W_j$;

- let $k_{ij}$ be the number of nodes of $N_G(v_j)$ that occur before $v_i$ in $C_i$;

- let $X_t^{ij}$ be the sum of a $t$-subset of $W_j \setminus \{W(v_i, v_j)\}$ drawn uniformly at random from the set of all such possible $t$-subsets; and finally

- let $Y_m^{ij}$ be the event $\{k_{ij} = m \ \wedge \ v_j \notin C_i\}$.

Then:

$$E[B_{v_i, v_j}] = \sum_{m=0}^{deg_G(v_j)-1} \Pr(Y_m^{ij}) \Pr\{X_m^{ij} \in [W_{cutoff}(v_j) - W(v_i, v_j), W_{cutoff}(v_j))\},$$

where $\Pr(Y_m^{ij})$ is obtained from Proposition 2:

$$\Pr(Y_m^{ij}) = \binom{deg_G(v_j) - 1}{m} \frac{m! \ (deg_G(v_j) - m)!}{(deg_G(v_j) + 1)!} = \frac{deg_G(v_j) - m}{deg_G(v_j)(deg_G(v_j) + 1)}.$$

Evaluating $\Pr\{X_m^{ij} \in [W_{cutoff}(v_j) - W(v_i, v_j), W_{cutoff}(v_j))\}$ is much more difficult because the distribution of $X_m^{ij}$ is a complicated function of the $deg_G(v_j) - 1$ numbers in $W_j \setminus \{W(v_i, v_j)\}$. However, we can obtain analytical expressions for the mean $\mu(X_m^{ij})$ and variance $\sigma^2(X_m^{ij})$. These are given by:

$$\mu(X_m^{ij}) = \frac{m}{deg_G(v_j) - 1} \ (\alpha_j - W(v_i, v_j))$$

$$\sigma^2(X_m^{ij}) = \frac{m(deg_G(v_j) - 1 - m)}{(deg_G(v_j) - 1)(deg_G(v_j) - 2)} \ (\beta_j - W(v_i, v_j)^2 - \frac{(\alpha_j - W(v_i, v_j))^2}{deg_G(v_j) - 1}).$$

Knowing only the mean and variance (not the exact distribution) of $X_m^{ij}$, we propose the approximation:

$$X_m^{ij} \sim \mathcal{N}(\mu(X_m^{ij}), \sigma^2(X_m^{ij})),$$

where $\mathcal{N}(\mu, \sigma^2)$ denotes the Gaussian random variable with mean $\mu$ and variance $\sigma^2$. This approximation is similar to the randomised approach that has been proposed and tested by Fatima et al. (2007).

With this approximation, we have:

$$Z_m^{ij} = \Pr\{X_m^{ij} \in [W_{cutoff}(v_j) - W(v_i, v_j), W_{cutoff}(v_j))\}$$

given by

$$Z_m^{ij} \approx \frac{1}{2}\left[\mathrm{erf}\left(\frac{W_{cutoff}(v_j) - \mu(X_m^{ij})}{\sqrt{2}\sigma(X_m^{ij})}\right) - \mathrm{erf}\left(\frac{W_{cutoff}(v_j) - W(v_i, v_j) - \mu(X_m^{ij})}{\sqrt{2}\sigma(X_m^{ij})}\right)\right].$$

This allows us to write:

$$E[B_{v_i, v_j}] = \sum_{m=0}^{deg_G(v_j)-1} \frac{deg_G(v_j) - m}{deg_G(v_j)(deg_G(v_j) + 1)} \, Z_m^{ij}.$$

The above equations are true only for $v_j \neq v_i$. For $v_j = v_i$ we have:

$$E[B_{v_i, v_i}] \approx \frac{1}{1 + deg_G(v_i)} \sum_{m=0}^{deg_G(v_i)} \Pr\{\mathcal{N}(\mu(X_m^{ii}), \sigma^2(X_m^{ii})) < W_{cutoff}(v_i)\},$$

where

$$\mu(X_m^{ii}) = \frac{m}{deg_G(v_i)}\alpha_i$$

and

$$\sigma^2(X_m^{ii}) = \frac{m \, (deg_G(v_i) - m)}{deg_G(v_i) \, (deg_G(v_i) - 1)} \, (\beta_i - \frac{\alpha_i^2}{deg_G(v_i)}).$$

Finally the Shapley value of node $v_i$ is given by $\sum_{v_j \in \{v_i\} \cup N_G(v_i)} E[B_{v_i, v_j}]$.

While in each graph it holds that $\sum_{v_i \in V(G)} deg_G(v_i) \leq 2|E|$, Algorithm 5 implements an $O(|V| + \sum_{v_i \in V(G)} \sum_{v_j \in N_G(v_i)} deg_G(v_j)) \leq O(|V| + |V||E|) = O(|V||E|)$ solution to compute the Shapley value for all agents in game $g_5$ using the above approximation.

Furthermore, we make the following observation: the approximation of the discrete random variable $X_m^{ij}$ as a continuous Gaussian random variable is good only when $deg_G(v_j)$ is large. For small $deg_G(v_j)$, one might as well use the brute force computation to determine $E[B_{v_i, v_j}]$ in $O(2^{deg_G(v_j)-1})$ time.

As far as directed graphs are concerned, in all calculations in Algorithm 5 we have to consider the indegree of a node instead of degree. Furthermore, the set of neighbours of a node $u$ should be defined as the set of nodes $v_i$ connected with directed edge $(u, v_i)$.

## 5. Simulations

In this section we evaluate the time performance of our exact algorithms for games $g_1$ to $g_4$ and our approximation algorithm for game $g_5$. In more detail, we compare our exact algorithms to the method of approximating the Shapley value via Monte Carlo sampling which has been the only feasible approach to compute game-theoretic network centrality available to date in the literature. First, we provide a detailed description of the simulation setup; then, we present data sets and the simulation results.

### 5.1 Simulation Setup

There are a few approximation methods for the Shapley value that have been recently proposed in literature. They can be divided into three groups—each referring to a specific subclass of coalitional games under consideration:

1. First, let us consider the method proposed by Fatima et al. (2007) and elaborated further by Fatima, Wooldridge, and Jennings (2008). This approach concerns *weighted voting games*. In these games, each player has a certain number of votes (or in other words, a weight). A coalition is "winning" if the number of votes in this coalition exceeds some specific threshold, or "losing" otherwise. Fatima et al. propose the following method to approximate the Shapley value in weighted voting games. Instead of finding marginal contributions of players to all $2^n$ coalitions, the authors consider only $n$ randomly-selected coalitions, one of each size (i.e., from 1 to $n$). Only for these $n$ coalitions are the player's expected marginal contributions calculated and the average of these contributions yields an approximation of the Shapley value. Whereas Fatima et al. method is certainly attractive, it is only applicable to games in which the value of a coalition depends on the sum of associated weights being in some bounds. This is not the case for our games $g_1$ to $g_4$.[11]

2. Another method was proposed by Bachrach, Markakis, Procaccia, Rosenschein, and Saberi (2008a) in the context of *simple coalitional games* [12] in which the characteristic function is binary—i.e., each coalition has a value of either zero or one. For these games, Bachrach et al. extend the approach suggested by Mann and Shapley (1960) and provide more rigorous statistical analysis. In particular, Mann and Shapley described the Monte Carlo simulations to estimate the Shapley value from a random sample of coalitions. Bachrach at al. use this technique to compute the Banzhaf power index and then they suggested using a random sample of permutations of all players in order to compute the Shapley-Shubik index for simple coalitional games.[13] The computation of the confidence interval, which is crucial in such an approach, hinges upon the binary form of the characteristic function for simple coalitional games. This

---

11. Recall that our approximation algorithm for $g_5$ builds upon Fatima et al. method. This is because in this game the marginal contribution of each node depends on the weights assigned to its incident edges.
12. Note that weighted voting games are simple coalitional games.
13. The Shapley-Shubik index is a very well-known application of the Shapley value that evaluates the power of individuals in voting (Shapley & Shubik, 1954).

---

**Algorithm 6:** Monte Carlo method to approximate the Shapley value

---

**Input**:

◇ Characteristic function $v$, maximum iteration $maxIter$

**Output**: Aproximation of Shapley value for game $v$

**for** $v_i \in V(G)$ **do**

   |   SV$[v_i] = 0$ ;

**end**

**for** $i = 1$ *to maxIter* **do**

   |   shuffle($V(G)$);

   |   ***Marginal Contribution*** **block**

   |    |   P $= \emptyset$ ;

   |    |   **for** $v_i \in V(G)$ **do**

   |    |    |   SV$[v_i]$ += $v$(P $\cup \{v_i\}$) - $v$(P) ;

   |    |    |   P $=$ P $\cup \{v_i\}$ ;

   |    |   **end**

**end**

**for** $v_i \in V(G)$ **do**

   |   SV$[v_i] = \frac{\text{SV}[v_i]}{maxIter}$ ;

**end**

return SV ;

---

method is more general than the one proposed by Fatima et al. (2007)—as weighted voting games are a subset of simple coalitional games—but still it cannot be effectively used for our games $g_1$ to $g_4$, where the characteristic functions are not binary.

3. Unlike the first two methods, the last method described by Castro et al. (2009) can be efficiently applied to all coalitional games in characteristic function game form, assuming that the worth of every coalition can be computed in polynomial time. Here, approximating the Shapley value involves generating permutations of all players and computing the marginal contribution of each player to the set of players occurring before it. The solution precision increases (statistically) with every new permutation analysed. Furthermore, the authors show how to estimate the appropriate size of a permutation sample in order to guarantee a low error. Given its broad applicability, this method is used in our simulations as a comparison benchmark.

In more detail, in a preliminary step, we test what is the maximum number of Monte Carlo iterations that can be performed in a reasonable time for any given game. This maximum number of iterations, denoted $maxIter$, becomes an input to Algorithm 6 for Monte Carlo sampling. In this algorithm, in each one of the $maxIter$ iterations, a random permutation of all nodes is generated. Then, using a characteristic function from the set $\nu \in \{\nu_1, \nu_2, \nu_3, \nu_4, \nu_5\}$, it calculates the marginal contribution of each node to the set $P$ of nodes occurring before a given node in a random permutation.[14] Finally, the algorithm divides the aggregated sum of all contributions for each node by the number of iterations performed. The time complexity of this algorithm is $O(maxIter * con)$, where $con$ denotes the number of operations necessary for computing the *Marginal Contribution* block. This block is specifically tailored to the particular form of the characteristic function of each of the games $g_1$ to $g_5$. In particular, for game $g_1$ (see Algorithm 6), it is constructed as follows. Recall that, in this game, node $v_i$ makes a positive contribution to coalition $P$ through itself and through some adjacent node $u$ under two conditions. Firstly, neither $v_i$ nor $u$ are in $P$. Secondly, there is no edge from $P$ to $v_i$ or $u$. To check for these conditions in Algorithm 6 we store those nodes that have already contributed to the value of coalition $P$ in an array called: *Counted*. For each node $v_i$, the algorithm iterates through the set of its neighbours and for each adjacent node it checks whether this adjacent node is counted in the array *Counted*. If not, the marginal contribution of the node $v_i$ is increased by one. In Appendix A we describe the *Marginal Contribution* block for games $g_2, \ldots, g_5$, respectively.[15]

Some details of how Algorithm 6 is applied to generate the Shapley value approximations for games $g_1$ to $g_4$, for which we propose exact polynomial solutions, differ from $g_5$, for which we developed an approximate solution. Specifically, for games $g_1$ to $g_4$:

1. We use the exact algorithm proposed in this paper to compute the Shapley value.

2. Then, we run Monte Carlo simulations 30 times.[16] In every run:

   - We perform $maxIter$ Monte Carlo iterations.
   - After every five iterations, we compare the approximation of the Shapley value obtained via Monte Carlo simulation with the exact Shapley value obtained with our algorithm.
   - We record the algorithm's runtime and the error, where the error is defined as the maximum discrepancy between the actual Shapley value and the Monte Carlo-based approximation of the Shapley value.

3. Finally, we compute the confidence interval using all iterations (0.95% confidence level).[17]

In the case of game $g_5$ we cannot determine the exact Shapley value for larger networks. Therefore, we performed two levels of simulation: one level on small networks and one level on large networks. Specifically:

1. For small networks, we generate 30 random instances of weighted complete graphs with 6 nodes (denoted $K_6$) and the same number of graphs with 12 nodes (denoted $K_{12}$) with weights drawn from a uniform distribution $U(0, 1)$. Then, for each graph and each of the two parameters $W_{cutoff}(v_i) = \frac{1}{4}\alpha(v_i)$ and $W_{cutoff}(v_i) = \frac{3}{4}\alpha(v_i)$:[18]

---

14. Recall that the characteristic functions $v_1, v_2, \ldots, v_5$ correspond to games $g_1, g_2, \ldots, g_5$, respectively.

15. The software package in C++ containing all our exact/approximation algorithms, as well as the Monte Carlo approximation algorithms are available at `www.tomaszmichalak.net`.

16. For the purpose of comparison to our method, it suffices to use 30 iterations, as the standard errors converge significantly to indicate the magnitude of the cost of using the Monte Carlo method.

17. Since for $g_4$ each Monte Carlo iteration is relatively time consuming, we run it only once; thus, no confidence interval is generated, i.e., the third step is omitted.

- We compute the exact Shapley value using formula (3).

- Then, we run our approximation algorithm and determine the error in our approximation.

- Finally, we run 2000 and 6000 Monte Carlo iterations for $K_6$ and $K_{12}$, respectively.

2. For large networks, we again generate 30 random instances of weighted complete graphs, but now with 1000 nodes (we denote them $K_{1000}$). Then, for each graph and each of the three parameters $W_{cutoff}(v_i) = \frac{1}{4}\alpha(v_i)$, $W_{cutoff}(v_i) = \frac{2}{4}\alpha(v_i)$, and $W_{cutoff}(v_i) = \frac{3}{4}\alpha(v_i)$:

- We run our approximation algorithm for the Shapley value.

- Then, we run the fixed number (200000) of Monte Carlo iterations.

- Finally, we compute how the Monte Carlo solution converges to the results of our approximation algorithm.

Having described the simulation setup, we will now discuss the data sets and, finally, the simulation results.

## 5.2 Data Used in Simulations

We consider two networks that have already been well-studied in the literature. Specifically, for games $g_1 - g_3$ we present simulations on an undirected, unweighted network representing the topology of the Western States Power Grid (WSPG).[19] This network (which has 4940 nodes and 6594 edges) has been studied in many contexts before (see, for instance, Watts & Strogatz, 1998) and is freely available online (see, e.g., `http://networkdata.ics.uci.edu/ data.php?id=107`). For games $g_3 - g_5$ (played on weighted networks), we used the network of astrophysics collaborations (abbreviated henceforth APhC) between Jan 1, 1995 and December 31, 1999. This network (which has 16705 nodes and 121251 edges) is also freely available online (see, e.g., `http://networkdata.ics.uci.edu/ data.php?id=13`) and has been used in previous studies like Newman (2001).

## 5.3 Simulation Results

The results presented in this section show that our exact algorithms are, in general, much faster then the Monte Carlo sampling, and this is the case even if we allow for generous error tolerance. Furthermore, requiring smaller Monte Carlo errors makes the Monte Carlo runtime exponentially slower than our exact solution.

---

18. Recall that $\alpha_j$ is the sum of all the weights in $W_j$ as defined in Section 4.5.

19. Note that with the distance threshold $d_{cutoff}$ replaced with a hop threshold $k_{cutoff}$, game $g_3$ can be played on an unweighted network.

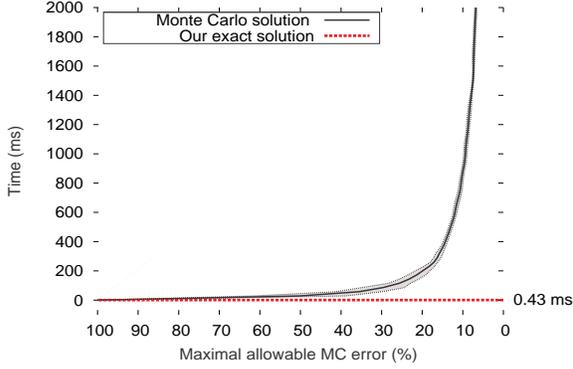

Figure 5: $g_1$, WSPG (UW)

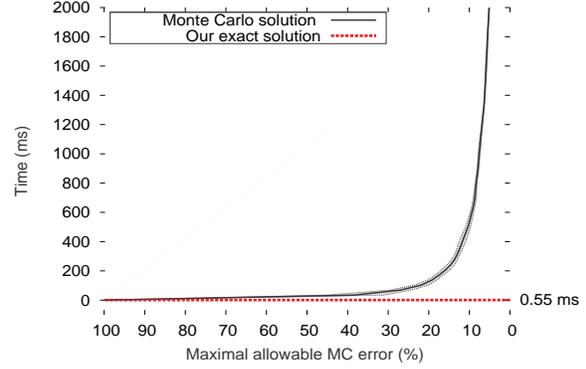

Figure 6: $g_2$, $k = 2$, WSPG (UW)

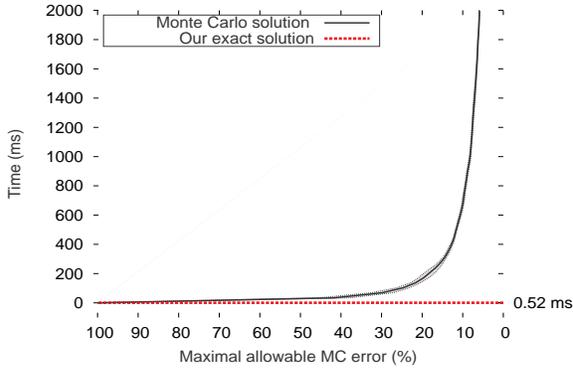

Figure 7: $g_2$, $k_i = \frac{deg_i}{2}$, WSPG (UW)

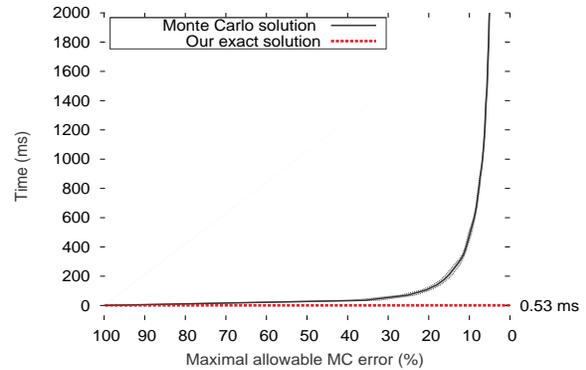

Figure 8: $g_2$, $k_i = \frac{3}{4} deg_i$, WSPG (UW)

In more detail, the simulation results for game $g_1$ are shown in Figure 5. The dotted line shows the performance of our exact algorithm which needs $0.43ms$ to compute the Shapley value. In contrast, generating any reasonable Monte Carlo result takes a substantially longer time (the solid line shows the average and the shaded area depicts the confidence interval for Monte Carlo simulations). In particular, it takes on average more than $200ms$ to achieve a 20% error and more than $2000ms$ are required to guarantee a 5% error (which is more than 4600 times slower than our exact algorithm).

Figures 6 - 8 concern game $g_2$ for different values of $k$ ($k = 2$, $k_i = \frac{deg_i}{2}$, and $k_i = \frac{3}{4} deg_i$, respectively, where $deg_i$ is the degree of node $v_i$).[20] The advantage of our exact algorithm over Monte Carlo simulation is again exponential.

Replacing the distance threshold $d_{cutoff}$ with a hop threshold $k_{cutoff}$ enables game $g_3$ to be played on an unweighted network. Thus, similarly to games $g_1$ and $g_2$, we test it on the Western States Power Grid. The results are shown on Figures 9 and 10 for $k_{cutoff}$ being equal to 2 and 3, respectively. The third game is clearly more computationally challenging than $g_1$ and $g_2$ (note that the vertical axis is in seconds instead of milliseconds). Now, our exact algorithm takes about $13s$ to complete. The much lower speedups of the exact methods with respect to Monte Carlo approach stem from the fact that both algorithms have to start with Dijkstra's algorithm. Although this algorithm has to be run only once in both cases it takes more than $12.5s$ for the considered network. This means that the exact solution is slower by orders of magnitude (compared to games $g_1$ and $g_2$). The Monte Carlo approach is also slower, but this slowdown is much less significant in relative terms.

---

20. Recall that in $g_2$ the meaning of parameter $k$ is as follows: the value of coalition $C$ depends on the number of nodes in the network with at least $k$ neighbours in $C$.

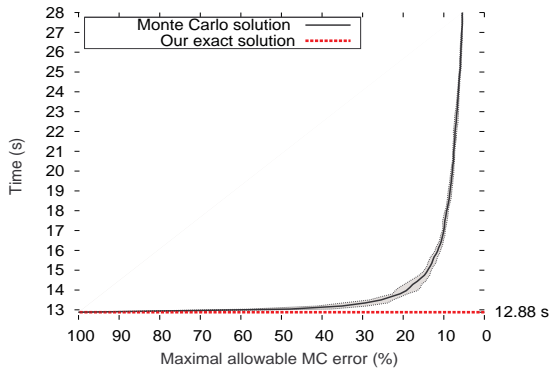 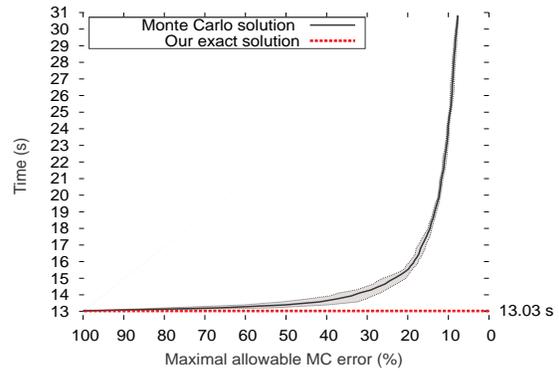

Figure 9: $g_3$, $k_{cutoff} = 2$, WSPG (UW)    Figure 10: $g_3$, $k_{cutoff} = 3$, WSPG (UW)

Figures 11 and 12 show the performance of the algorithms for game $g_3$ on the astrophysics collaboration network that, unlike the Western States Power Grid, is a weighted network. We observe that increasing the value of $d_{cutoff}$ (here from $d_{cutoff} = \frac{d_{avg}}{8}$ to $d_{cutoff} = \frac{d_{avg}}{4}$) significantly worsens the performance of the Monte Carlo-based algorithm. This is because the increasing number of nodes that have to be taken into account while computing marginal contributions (see the inner loop in Algorithm 8) is not only more time consuming, but also increases the Monte Carlo error.

For game $g_4$ the performance of algorithms is shown in Figures 13 - 15 (for $f(d) = \frac{1}{1+d}$, $f(d) = \frac{1}{1+d^2}$ and $f(d) = e^{-d}$, respectively). Whereas the Monte Carlo methods for the first three games are able to achieve a reasonable error bound in seconds or minutes, for the fourth game it takes more than 40 hours to approach 50% error. This is because the inner loop of the *Marginal Contribution* block (see Algorithm 9) iterates over all nodes in the network. Due to the time consuming performance we run the simulations only once. Interestingly, we observe that the error of the Monte Carlo method sometimes increases slightly when more iterations are performed. This confirms that the error of the Monte Carlo method to approximate the Shapley value proposed in Castro et al. (2009) is only statistically decreasing in time. Certain new randomly chosen permutations can actually increase the error.

Figures 16, 17, 18 and 19 present comparisons of our approximation algorithm for game $g_5$ against Monte Carlo sampling for small networks (for which the exact Shapley value can be computed from the definition in formula (3)). In these figures, the horizontal dotted line shows the running time of our solution, while the vertical dotted line shows its average approximation error with the shaded area being the confidence interval. As previously, the

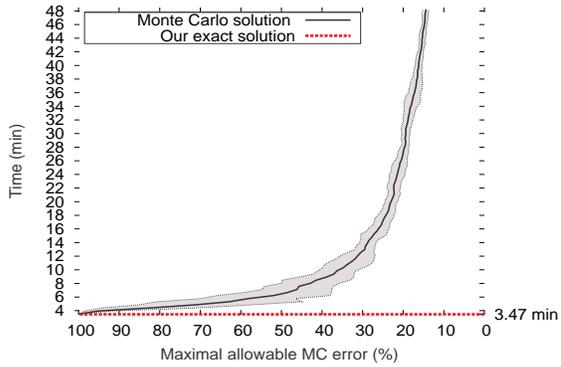

Figure 11: $g_3$, $d_{cutoff} = \frac{d_{avg}}{8}$, APhC (W)

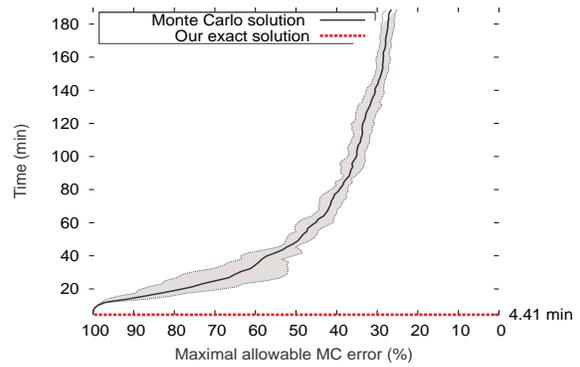

Figure 12: $g_3$, $d_{cutoff} = \frac{d_{avg}}{4}$, APhC (W)

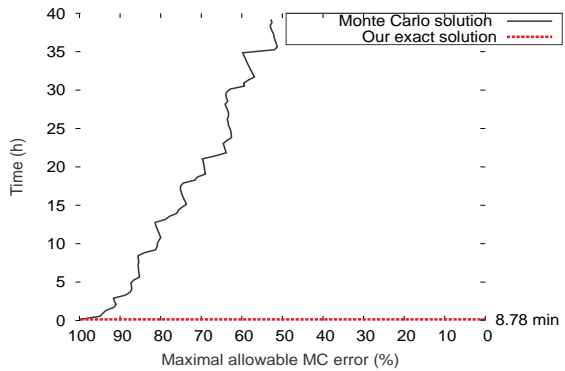

Figure 13: $g_4$, $f(d) = \frac{1}{1+d}$, APhC (W)

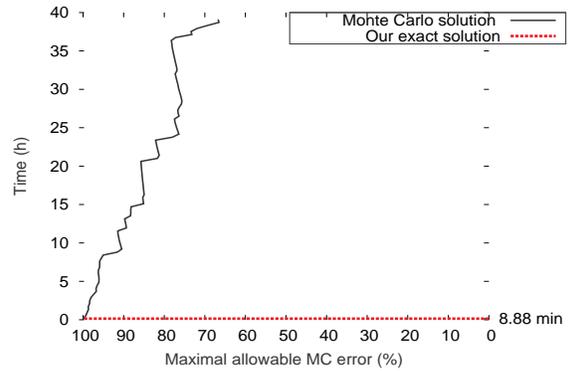

Figure 14: $g_4$, $f(d) = \frac{1}{1+d^2}$, APhC (W)

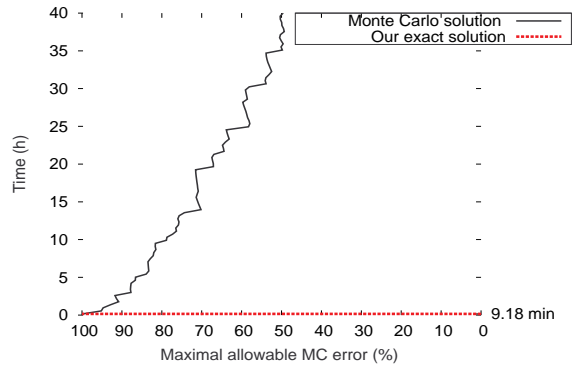

Figure 15: $g_4$, $f(d) = e^{-d}$, APhC (W)

solid line shows the average, and the shaded area depicts the confidence interval for the Monte Carlo simulations. We see in Figures 16, 17 and 18 that the approximation error in our proposed algorithm is well-contained for small networks. Specifically, for $K_6$ it is about 10%; whereas for the bigger network $K_{12}$ it is about 5%. However, we notice that, for higher values of $W_{cutoff}$, the Monte Carlo method may slightly outperform our solution. See in Figure 17 how the average approximation error of the Monte Carlo sampling achieved in $0.38ms$ is lower than the average error achieved by our method. Already for $K_{12}$ this effect does not occur (see Figure 19).

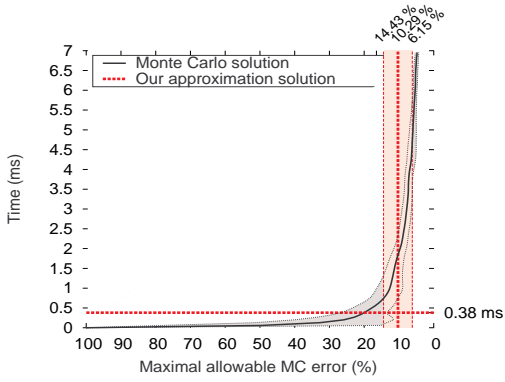

Figure 16: $g_5$, $W_{cutoff} = \frac{1}{4}\alpha_i$, $K_6$ (W)

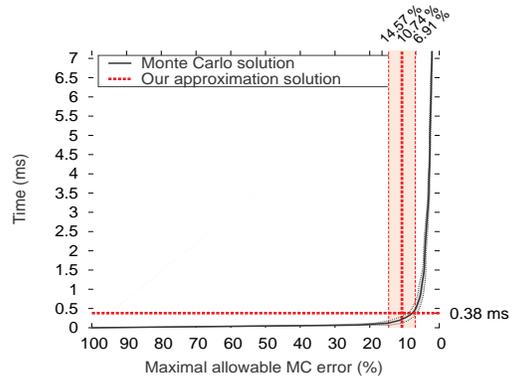

Figure 17: $g_5$, $W_{cutoff} = \frac{3}{4}\alpha_i$, $K_6$ (W),

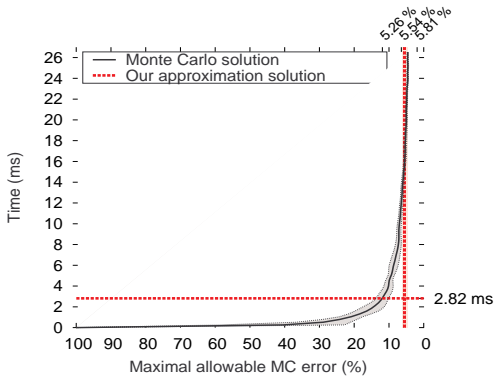

Figure 18: $g_5$, $W_{cutoff} = \frac{1}{4}\alpha_i$, $K_{12}$ (W)

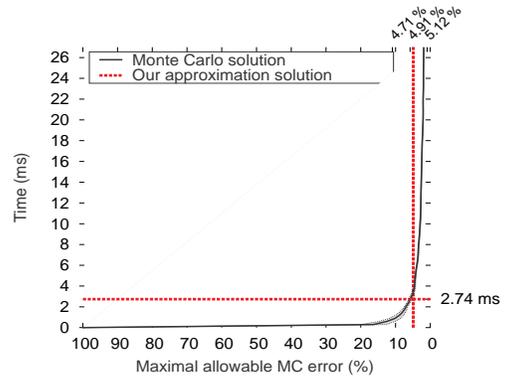

Figure 19: $g_5$, $W_{cutoff} = \frac{3}{4}\alpha_i$, $K_{12}$ (W)

For large networks, where the exact Shapley value cannot be obtained, we are naturally unable to compute exact approximation error. We believe that this error may be higher than the values obtained for $K_6$ and $K_{12}$. However, the mixed strategy, that we discussed in Section 4 and that uses our approximation only for large degree vertices, should work towards containing the error within practical tolerance bounds. As far as we believe that Monte Carlo gives good results, from Figure 20, we can infer that our approximation solution for large networks gives good results (within 5%) and is at least two times faster than the Monte Carlo algorithm.

To summarise, our exact solutions outperform Monte Carlo simulations even if relatively wide error margins are allowed. However, this is not always the case for our approximation algorithm for game $g_5$. Furthermore, it should be underlined that if the centrality metrics under consideration cannot be described with any of the games $g_1$ to $g_4$ for which exact algorithms are now available, then Monte Carlo simulations are still a viable option.

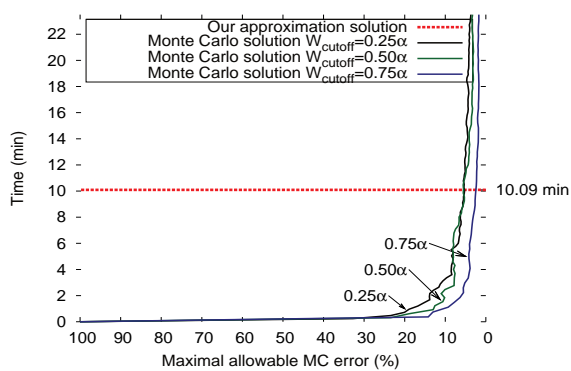

Figure 20: $g_5$, $W_{cutoff} = \frac{1}{4}\alpha_i$, $W_{cutoff} = \frac{2}{4}\alpha_i$, and $W_{cutoff} = \frac{3}{4}\alpha_i$, $K_{1000}$ (W)

## 6. Conclusions and Future Work

The key finding of this paper is that the Shapley value for many centrality-related cooperative games of interest played on networks can be solved analytically. The resulting algorithms are not only error-free, but also run in polynomial time and, in practice, are much faster than Monte Carlo methods. Approximate closed-form expressions and algorithms can also be constructed for some classes of games played on weighted networks. Simulation results show that these approximations are acceptable for a range of situations.

There are a number of directions for future work. On one hand, the Shapley value-based extensions of other centrality notions, that suit particular applications, can be developed. As a step in this direction, the first study of the Shapley value-based betweenness centrality has been recently presented by Szczepański, Michalak, and Rahwan (2012). On the other hand, it would be interesting to analyze what other coalitional games defined over a network would better reflect centrality of nodes in certain real-life applications. In this spirit, recent works of del Pozo, Manuel, González-Arangüena, and Owen (2011) and Amer, Giménez, and Magana (2012) focus on generalized coalitional games in which the order of agents forming coalitions matter. Nevertheless, there are still other classes of coalitional games, such as games with either positive or negative externalities (Yi, 1997), that have been extensively studied in game theory and that may yield interesting results when applied to network centrality. Another interesting application for which a new class of coalitional games defined over a network could be developed is the problem of influence maximization, already mentioned in the introduction.

It is also interesting to analyse the properties of game-theoretic network centralities constructed on solution concepts from cooperative game theory other than the Shapley value. In particular, if the game defined over a network belongs to the class of simple coalitional games (i.e., with a binary characteristic function) then the Banzhaf power index (Banzhaf, 1965) could be also used as a centrality metric. Otherwise, more general solution concepts such as the core (Osborne & Rubinstein, 1994) or the nucleolus (Schmeidler, 1969) could be applied.

Ultimately, it would be interesting to develop a more formal and general approach that would allow us to construct coalitional games defined over networks that correspond to other known centrality metrics or even entire families of them.[21] Such an approach would involve developing a group centrality first and then building a characteristic function of a coalitional game upon it. Of course, while developing new centrality metrics based on coalitional games, one should keep in mind computational properties of the proposed solutions. Although we were able to obtain satisfactory computational results for the games considered in this paper, the computation of the game-theoretic network centrality may become much more challenging for more complex definitions of the characteristic function.

## Acknowledgments

We would like to thank three anonymous reviewers for their comments on the earlier version of the paper that helped to improve it considerably. Also, we would like to thank dr Talal Rahwan and dr Suri Rama Narayanam for proofreading, helpful comments and suggestions. Tomasz Michalak was partially supported by the European Research Council under Advanced Grant 291528 ("RACE"). Nicholas R. Jennings (and partially Tomasz Michalak) was supported by the ORCHID Project, funded by EPSRC (Engineering and Physical Research Council) under the grant EP/I011587/1.

## Appendix A. Marginal Contribution Blocks for Algorithm 6 for $g_2$-$g_5$

---

**Algorithm 7:** ***Marginal Contribution*** block of Algorithm 6 for $g_2$

---

Counted $\leftarrow$ false ;
Edges $\leftarrow 0$ ;
**foreach** $v_i \in V(G)$ **do**
    **foreach** $u \in N_G(v_i) \cup \{v_i\}$ **do**
        Edges[u]$++$ ;
        **if** *!Counted[u] and ( Edges[u] $\geq$ k[u] or u = $v_i$ )* **then**
            SV[$v_i$]$++$ ;
            Counted[$u$] = true ;
        **end**
    **end**
**end**

---

For each of the games considered in our paper the ***Marginal Contribution*** block of Algorithm 6 takes a slightly different form. In the main text we explained the functioning of this block for $g_1$. In this appendix, we discuss this block for the remaining four games. In particular:

$g_2$: Here, node $v_i$ makes a positive contribution to a coalition $P$ both through itself and through some adjacent node $u$ also under two conditions. Firstly, neither $v_i$ nor $u$ are

---

21. We thank an anonymous reviewer for this suggestion.

---

**Algorithm 8:** *Marginal Contribution* block of Algorithm 6 for $g_3$

---

Counted ← false ;
**foreach** $v_i \in V(G)$ **do**
    **foreach** $u \in extNeighbors(v_i) \cup \{v_i\}$ **do**
        **if** *!Counted[u]* **then**
            SV[$v_i$]++ ;
            Counted[u] = true ;
        **end**
    **end**
**end**

---

**Algorithm 9:** *Marginal Contribution* block of Algorithm 6 for $g_4$

---

dist ← infinity ;
**foreach** $v_i \in V(G)$ **do**
    **foreach** $u \in V(G)$ **do**
        **if** *D[u] < dist[u]* **then**
            SV[$v_i$] += f(D[u]) - f(dist[u]) ;
            dist[u] = D[u] ;
        **end**
    **end**
    SV[$v_i$] += f(dist[u]) - f(0) ;
    dist[$v_i$] = 0 ;
**end**

---

in $P$. Secondly, there is less than $k$ edges from $P$ to $v_i$ and there is exactly $k-1$ edges from $P$ to $u$. In order to check the first condition in Algorithm 7 we use the array *Counted*, and to check the second one, we use the array *Edges*. For each node $v_i$, the algorithm iterates through the set of its neighbours and for each adjacent node it checks whether this adjacent node meets these two conditions. If so, then the marginal contribution of the node $v_i$ is increased by one.

$g_{3/4}$: In *Marginal Contribution* blocks for games $g_3$ and $g_4$ (Algorithms 8 and 9), all the values that are dependent on the distance (*extNeighbours* and $D$) are calculated using Dijkstra's algorithm and stored in memory. These pre-computations allow us to significantly speed up Monte Carlo methods. Now, in $g_3$ node $v_i$ makes a positive contribution to coalition $P$ through itself and through some adjacent node $u$ under two conditions. Firstly, neither $v_i$ nor $u$ are in $P$. Secondly, there is no edge length of $d_{cutoff}$ from $P$ to $v_i$ or $u$. To check for these conditions in Algorithm 8 we again use the array *Counted*. For each node $v_i$, the algorithm iterates through the set of its extended neighbours and for each of them it checks whether this neighbour meets the conditions. If so, the marginal contribution of the node $v_i$ is increased by one. In game $g_4$, node $v_i$ makes a positive contribution to coalition $P$ through each node (including itself) that is closer to $v_i$ than to $P$. In Algorithm 9 we use array *Dist* to store distances

---

**Algorithm 10:** *__Marginal Contribution__* block of Algorithm 6 for $g_5$

---

Counted ← false ;
Weights ← 0 ;
**foreach** $v_i \in V(G)$ **do**
 **foreach** $u \in N_G(v_i) \cup \{v_i\}$ **do**
  weights[u]+= $W(v_i, u)$;
  **if** *!Counted[u] and ( weights[u] $\geq W_{cutoff}(u)$ or $u = v_i$ )* **then**
   SV[$v_i$]++ ;
   Counted[u] = true ;
  **end**
 **end**
**end**

---

from coalition $P$ to all nodes in the graph and array $D$ to store all distances from $v_i$ to all other nodes. For each node $v_i$, the algorithm iterates through all nodes in the graph, and for each node $u$, if the distance from $v_i$ to $u$ is smaller than from $P$ to $u$, the algorithm computes the marginal contribution as $f(D[u]) - f(Dist[u])$. The value $Dist[u]$ is then updated to $D[u]$—this is a new distance from $P$ to $u$.

$g_5$: In game $g_5$, which is an extension of $g_2$ to weighted graphs, node $v_i$ makes a positive contribution to coalition $P$ (both through itself and through some adjacent node $u$) under two conditions. Firstly, neither $v_i$ nor $u$ are in $P$. Secondly, the sum of weights on edges from $P$ to $v_i$ is less than $W_{cutoff}(v_i)$ and the sum of weights on edges from $P$ to $u$ is greater than, or equal to, $W_{cutoff}(u) - W(v_i, u)$ and smaller than $W_{cutoff}(v_i) + W(v_i, u)$. In order to check the first condition in Algorithm 10 we use the array *Counted*, and to check the second one, we use the array *Weights*. For each node $v_i$, the algorithm iterates through the set of its neighbours and for each adjacent node it checks whether this adjacent node meets these two conditions. If so, then the marginal contribution of the node $v_i$ is increased by one.

## Appendix B: Main Notation Used in the Paper

| | |
|---|---|
| $A$ | The set of players. |
| $a_i$ | A player in $A$. |
| $C$ | A coalition. |
| $\nu(C)$ | A value of the coalition, where $\nu$ is characteristic function. |
| $(A, \nu)/g_i$ | A coalitional game. |
| $SV_{g_j}(v_i)$ | The Shapley value od the vertex $v_i$ in game $g_j$. |
| $G = (V, E)$ | Unweighted graph/network consisting of the set of vertices $V$ and edges $E$. |
| $G = (V, E, W)$ | Weighted graph/network. |
| $W(v, u)$ | Weight on the edge from $v$ to $u$. |
| $V(G)/V, E(G)/E$ | the set of vertices and edges in graph $G$. |
| $v_i \in V$ | The vertex from the set $V$. |
| $deg(v_i)$ | Degree of the vertex $v_i$. |
| $N_G(v_i)$ | Set of neighbours of vertex $v_i \in G$. |
| $distance(v, u)/d(v, u)$ | The distance between vertices $v$ and $u$. |
| $N_G(v_i, d_{cutoff})$ | Extended neighbourhood: $N_G(v_j, d_{cutoff}) = \{v_k \neq v_j : \text{distance}(v_k, v_j) \leq d_{cutoff}\}$. |
| $MC(u, v)$ | Marginal contribution that vertex $u$ makes through vertex $v$. |
| $\Pi(A)$ | The set of all orders of players in $A$. |
| $\pi \in \Pi(A)$ | The single ordering of agents in $A$. |
| $\pi(i)$ | The position of $i - th$ element in ordering $\pi$. |
| $C_\pi(i)$ | $\{a_j \in \pi : \pi(j) < \pi(i)\}$. |
| $fringe(C)$ | $\{v \in V(G) : v \in C$ (or) $\exists u \in C$ such that $(u, v) \in E(G)\}$. |
| $k(v_i)/k_i$ | The number assigned to vertex $v$ used in Game 2. The minimum number of adjacent nodes necessary to influence node $v_i$. |
| $W_{cutoff}(v_i)$ | The number assigned to vertex $v$ used in Game 5. Minimum sum of weights on adjacent edges necessary to influence node $v_i$. |
| $\mathbb{E}[\cdot]$ | The expectation operator. |
| $\mathbb{P}[\cdot]$ | The probability operator. |
| $O(\cdot)$ | The big $O$ complexity notation. |
| $B, X, Y$ | Random variables. |
| $\mathcal{N}(\mu, \sigma^2)$ | Normal distribution with mean $\mu$ and variance $\sigma^2$. |
| $erf(\cdot)$ | The error function. |
| $\alpha_j$ | The sum of all the weights of incident edges to vertex $v_j$. |
| $\beta_j$ | The sum of the squares of all the weights of incident edges to vertex $v_j$. |
| $f(.)$ | A positive valued decreasing function. |
| $K_i$ | The complete graph (clique) with $i$ nodes. |

# References

Aadithya, K., Michalak, T., & Jennings, N. (2011). Representation of coalitional games with algebraic decision diagrams. In *AAMAS '11: Proceedings of the 10th International Joint Conference on Autonomous Agents and Multi-Agent Systems*, pp. 1121–1122.

Amer, R., Giménez, J., & Magana, A. (2012). Accessibility measures to nodes of directed graphs using solutions for generalized cooperative games. *Mathematical Methods of Operations Research, 75*, 105–134.

Aziz, H., Lachish, O., Paterson, M., & Savani, R. (2009a). Power indices in spanning connectivity games. In *AAIM '09: Proceedings of the 5th International Conference on Algorithmic Aspects in Information and Management*, pp. 55–67.

Aziz, H., Lachish, O., Paterson, M., & Savani, R. (2009b). Wiretapping a hidden network. In *WINE '09: Proceedings of the the 5th Workshop on Internet & Network Economics*, pp. 438–446.

Bachrach, Y., Markakis, E., Procaccia, A. D., Rosenschein, J. S., & Saberi, A. (2008a). Approximating power indices. In *AAMAS '08: Proceedings of the 7th International Joint Conference on Autonomous Agents and Multi-Agent Systems*, pp. 943–950.

Bachrach, Y., Rosenschein, J. S., & Porat, E. (2008b). Power and stability in connectivity games. In *AAMAS '08: Proceedings of the 7th International Joint Conference on Autonomous Agents and Multi-Agent Systems*, pp. 999–1006.

Bachrach, Y., & Rosenschein, J. (2009). Power in threshold network flow games. *Autonomous Agents and Multi-Agent Systems, 18*(1), 106–132.

Banzhaf, J. F. (1965). Weighted Voting Doesn't Work: A Mathematical Analysis. *Rutgers Law Rev., 19*, 317–343.

Bikhchandani, S., Hirshleifer, D., & Welch, I. (1992). A theory of fads, fashion, custom, and cultural change in informational cascades. *Journal of Political Economy, 100*(5), 992–1026.

Bolus, S. (2011). Power indices of simple games and vector-weighted majority games by means of binary decision diagrams. *European Journal of Operational Research, 210*(2), 258–272.

Bonacich, P. (1972). Factoring and weighting approaches to status scores and clique identification. *Journal of Mathematical Sociology, 2*(1), 113–120.

Bonacich, P. (1987). Power and centrality: A family of measures. *American Journal of Sociology, 92*(5), 1170–1182.

Borgatti, S. P., & Everett, M. (2006). A graph-theoretic framework for classifying centrality measures. social networks. *Social Networks, 28(4)*, 466–484.

Brandes, U. (2001). A faster algorithm for betweenness centrality. *Journal of Mathematical Sociology, 25*(2), 163–177.

Brandes, U., & Erlebach, T. (2005). *Network Analysis: Methodological Foundations*. Lecture notes in computer science: Tutorial. Springer.

Castro, J., Gomez, D., & Tejada, J. (2009). Polynomial calculation of the shapley value based on sampling. *Computers & Operations Research, 36*(5), 1726–1730.

Chalkiadakis, G., Elkind, E., & Wooldridge, M. (2011). *Computational Aspects of Cooperative Game Theory*. Synthesis Lectures on Artificial Intelligence and Machine Learning. Morgan & Claypool Publishers.

Conitzer, V., & Sandholm, T. (2004). Computing Shapley Values, manipulating value division schemes and checking core membership in multi-issue domains. In *AAAI '04: Proceedings of the Nineteenth National Conference on Artificial Intelligence*, pp. 219–225.

Cormen, T. (2001). *Introduction to algorithms*. MIT Press.

del Pozo, M., Manuel, C., González-Arangüena, E., & Owen, G. (2011). Centrality in directed social networks. a game theoretic approach. *Social Networks*, *33*(3), 191–200.

Deng, X., & Papadimitriou, C. (1994). On the complexity of cooperative solution concepts. *Mathematics of Operations Research*, *19*(2), 257–266.

Elkind, E., Goldberg, L., Goldberg, P., & Wooldridge, M. (2009). A tractable and expressive class of marginal contribution nets and its applications. *Mathematical Logic Quarterly*, *55*(4), 362–376.

Eppstein, D., & Wang, J. (2001). Fast approximation of centrality. In *SODA '01: Proceedings of the Twelfth Annual ACM-SIAM Symposium on Discrete Algorithms*, pp. 228–229.

Everett, M. G., & Borgatti, S. P. (1999). The centrality of groups and classes.. *Journal of Mathematical Sociology*, *23*(3), 181–201.

Fatima, S. S., Wooldridge, M., & Jennings, N. (2007). A randomized method for the shapley value for the voting game. In *AAMAS '07: Proceedings of the 11th International Joint Conference on Autonomous Agents and Multi-Agent Systems*, pp. 955–962.

Fatima, S. S., Wooldridge, M., & Jennings, N. (2008). A linear approximation method for the shapley value. *Artificial Intellilgence*, *172*(14), 1673–1699.

Freeman, L. (1979). Centrality in social networks: Conceptual clarification. *Social Networks*, *1*(3), 215–239.

Gómez, D., González-Arangüena, E., Manuel, C., Owen, G., Del Pozo, M., & Tejada, J. (2003). Centrality and power in social networks: A game theoretic approach. *Mathematical Social Sciences*, *46*(1), 27–54.

Goyal, A., Bonchi, F., & Lakshmanan, L. V. (2010). Learning influence probabilities in social networks. In *WSDM '10: Proceedings of the 3rd ACM international conference on Web search and data mining*, pp. 241–250.

Granovetter, M. (1978). Threshold models of collective behavior. *American Journal of Sociology*, *83*(6), 1420–1443.

Greco, G., Malizia, E., Palopoli, L., & Scarcello, F. (2009). On the complexity of compact coalitional games. In *IJCAI '09: Proceedings of the Twenty First International Joint Conference on Artifical Intelligence*, pp. 147–152.

Grofman, B., & Owen, G. (1982). A game-theoretic approach to measuring centrality in social networks. *Social Networks*, *4*, 213–224.

Ieong, S., & Shoham, Y. (2005). Marginal contribution nets: a compact representation scheme for coalitional games. In *EC '05: Proceedings of the Sixth ACM Conference on Electronic Commerce*, pp. 193–202.

Irwin, M., & Shapley, L. S. (1960). *Values of Large Games, IV : Evaluating the Electoral College by Montecarlo Techniques. CA* : RAND Corporation, Santa Monica.

Jeong, H., Mason, S. P., Barabasi, A. L., & Oltvai, Z. N. (2001). Lethality and centrality in protein networks. *Nature*, *411*(6833), 41–42.

Kempe, D., Kleinberg, J., & Tardos, É. (2003). Maximizing the spread of influence through a social network. In *KDD '03: Proceedings of the Ninth ACM SIGKDD International Conference on Knowledge Discovery and Data Mining*, pp. 137–146.

Kempe, D., Kleinberg, J., & Tardos, É. (2005). Influential nodes in a diffusion model for social networks. *Automata, Languages and Programming, 3580*, 99.

Koschützki, D., Lehmann, K., Peeters, L., Richter, S., Tenfelde-Podehl, D., & Zlotowski, O. (2005). *Centrality indices. Network analysis*, Vol. 3418 of *Lecture Notes in Computer Science*, pp. 16–61. Springer.

Lee, M.-J., Lee, J., Park, J. Y., Choi, R. H., & Chung, C.-W. (2012). Qube: a quick algorithm for updating betweenness centrality. In Mille, A., Gandon, F. L., Misselis, J., Rabinovich, M., & Staab, S. (Eds.), *WWW*, pp. 351–360. ACM.

Mann, I., & Shapley, L. (1962). Values of large games, VI: Evaluating the electoral college exactly.. RAND Research Memorandum.

Matsui, T., & Matsui, Y. (2000). A survey of algorithms for calculating power indices of weighted majority games. *Journal of the Operations Research Society of Japan, 43*, 71–86.

Michalak, T., Marciniak, D., Samotulski, M., Rahwan, T., McBurney, P., Wooldridge, M., & Jennings, N. (2010a). A logic-based representation for coalitional games with externalities. In *AAMAS '10: Proceedings of the 9th International Joint Conference on Autonomous Agents and Multi-Agent Systems*, pp. 125–132.

Michalak, T., Rahwan, T., Marciniak, D., Szamotulski, M., & Jennings, N. (2010b). Computational aspects of extending the Shapley Value to coalitional games with externalities. In *ECAI '10: Proceedings of the Nineteenth European Conference on Artificial Intelligence*.

Myerson, R. (1977). Graphs and cooperation in games. *Mathematics of Operations Research, 2*(3), 225–229.

Nagamochi, H., Zeng, D., Kabutoya, N., & Ibaraki, T. (1997). Complexity of the minimum base game on matroids. *Mathematics of Operations Research, 22*(1), 146–164.

Newman, M. (2001). Scientific collaboration networks. II. Shortest paths, weighted networks, and centrality. *Physical Review E, 64*(1), 016132(1–7).

Noh, J., & Rieger, H. (2004). Random walks on complex networks. *Physical Review Letters, 92*(11), 118701(1–4).

Osborne, M. J., & Rubinstein, A. (1994). *A Course in Game Theory*, Vol. 1 of *MIT Press Books*. The MIT Press.

Reka, A., & Barabási, A. (2002). Statistical mechanics of complex networks. *Reviews of Modern Physics, 74*, 47–97.

Sakurai, Y., Ueda, S., Iwasaki, A., Minato, S., & Yokoo, M. (2011). A compact representation scheme of coalitional games based on multi-terminal zero-suppressed binary decision diagrams. In *PRIMA '11: Public Risk Management Association*, pp. 4–18.

Schmeidler, D. (1969). The nucleolus of a characteristic function game. *SIAM Journal on Applied Mathematics*, 1163–1170.

Schultes, D., & Sanders, P. (2007). Dynamic highway-node routing. In *SEA '07: Proceedings of 6th Workshop on Experimental and Efficient Algorithms. LNCS*, pp. 66–79. Springer.

Shapley, L. S. (1953). A value for n-person games. In Kuhn, H., & Tucker, A. (Eds.), *In Contributions to the Theory of Games, volume II*, pp. 307–317. Princeton University Press.

Shapley, L. S., & Shubik, M. (1954). A Method for Evaluating the Distribution of Power in a Committee System. *The American Political Science Review*, *48*(3), 787–792.

Stephenson, K., & Zelen, M. (1989). Rethinking centrality: Methods and examples. *Social Networks*, *11*(1), 1–37.

Suri, N., & Narahari, Y. (2008). Determining the top-k nodes in social networks using the shapley value. In *AAMAS '08: Proceedings of the 7th International Joint Conference on Autonomous Agents and Multi-Agent Systems*, pp. 1509–1512.

Suri, N., & Narahari, Y. (2010). A Shapley Value based approach to discover influential nodes in social networks. *IEEE Transaction on Automation Science and Engineering*, *99*, 1–18.

Szczepański, P. L., Michalak, T., & Rahwan, T. (2012). A new approach to betweenness centrality based on the shapley value. In *AAMAS '12: Proceedings of the 11th International Joint Conference on Autonomous Agents and Multi-Agent Systems*, pp. 239–246.

Valente, T. (1996). Social network thresholds in the diffusion of innovations. *Social Networks*, *18*(1), 69–89.

Watts, D., & Strogatz, S. (1998). Collective dynamics of small-world networks. *Nature*, *393*(6684), 440–442.

Wooldridge, M., & Dunne, P. (2006). On the computational complexity of coalitional resource games. *Artificial Intelligence*, *170*(10), 835–871.

Yi, S.-S. (1997). Stable coalition structures with externalities. *Games and Economic Behavior*, *20*(2), 201–237.

Young, H. P. (2006). The Diffusion of Innovations in Social Networks. In Blume, B. L., & Durlauf, S. N. (Eds.), *Economy as an evolving complex system*, Vol. 3 of *Proceedings volume in the Santa Fe Institute studies in the sciences of complexity Santa Fe Institute Studies on the Sciences of Complexity*, pp. 267–282. Oxford University Press US.

Zolezzi, J. M., & Rudnick, H. (2002). Transmission cost allocation by cooperative games and coalition formation. *IEEE Transactions on Power Systems*, *17*, 1008–1015.